\documentclass[12pt]{article}

\usepackage{amssymb}
\usepackage{epsf}
\usepackage{rotating}

   \let\b=\beta   \let\g=\gamma   
         
      \let\l=\lambda  
\let\n=\nu

\def\slashed{\ds}

\def\b{\beta}

\def\l{\lambda}

\def\g{\gamma}

\def\ds#1{#1\kern-1ex\hbox{/}}
\def\dsh{h\kern-1.2ex /}

\newcommand{\bea}{\begin{eqnarray}}
\newcommand{\eea}{\end{eqnarray}}

\def\beq{\begin{equation}}
\def\eeq{\end{equation}}

\def\beqn{\begin{eqnarray}}
\def\eeqn{\end{eqnarray}}
\def\ba{\begin{eqnarray}}
\def\ea{\end{eqnarray}}

\def\slash#1{#1\hskip-6pt/\hskip6pt}

\newcommand{\beqa}{\begin{eqnarray}}
\newcommand{\eeqa}{\end{eqnarray}}

\newcommand{\la}{\lambda}

\begin{document}

\begin{center}
\vspace{1.5cm}
{\bf\large St\"{u}ckelberg Axions and the Effective Action of Anomalous Abelian Models.\\}
{\bf A $SU(3)_C\times SU(2)_W\times U(1)_Y\times U(1)_B$ model and its signature at the LHC}

\vspace{0.5cm}
{\bf\large Claudio Corian\`{o} $^{a}$  Nikos Irges $^{b}$ and Simone Morelli$^{a}$} 

\vspace{1cm}

{\it  $^a$Dipartimento di Fisica, Universit\`{a} del Salento \\
and  INFN Sezione di Lecce,  Via Arnesano 73100 Lecce, Italy}\\
\vspace{.5cm}
{\it and} \\

~\\
{\it
$^b$Department of Physics and Institute of Plasma Physics \\
 University of Crete, 71003 Heraklion, Greece\\}
\vspace{.12in}
~\\
\vspace{.12in}
\centerline{\em \bf Dedicated to the Memory of Hidenaga Yamagishi}
\begin{abstract}
We elaborate on an extension of the Standard Model with a gauge structure enlarged by a single 
anomalous $U(1)$, where the presence of a Wess-Zumino term is motivated 
by the Green-Schwarz mechanism of string theory. The additional gauge interaction 
is anomalous and requires an axion for anomaly cancelation. The pseudoscalar 
implements the St\"{u}ckelberg mechanism and undergoes mixing with the standard Higgs sector 
to render the additional $U(1)$ massive. We consider a 2-Higgs doublet model.
We show that the anomalous effective vertices involving neutral currents
are potentially observable. We clarify their role 
in the case of simple processes such as $Z^*\to \gamma \gamma$, which are at variance 
with respect to the Standard Model. A brief discussion of the implications of these studies for the LHC is included.

\end{abstract}
\end{center}
\newpage

\section{Introduction}

Among the possible extensions of the Standard Model (SM), those where the 
$SU(3)_C\times SU(2)_W\times U(1)_Y$ gauge group is enlarged by a number of
extra $U(1)$ symmetries are quite attractive for being modest enough departures 
from the SM so that they are computationally tractable, 
but at the same time predictive enough so that they are interesting and even 
perhaps testable at the LHC.
Of particular popularity among these have been models where at least one of the extra
$U(1)$'s is "anomalous", that is, some of the fermion triangle loops with
gauge boson external legs are non-vanishing. The existence of this possibility
was noticed in the context of the (compactified to four dimensions) 
heterotic superstring where the stability of the supersymmetric
vacuum \cite{DSW} can trigger in the four-dimensional low energy 
effective action a non-vanishing Fayet-Iliopoulos term proportional to the 
gravitational anomaly, i.e. proportional to the anomalous trace of the corresponding $U(1)$.  
The mechanism was recognized to be the low energy manifestation of the Green-Schwarz anomaly (GS)
cancellation mechanism of string theory.\footnote{Conventionally in this paper 
we will use both the term ``Green-Schwarz'' (GS) to denote the mechanism 
of cancelation of the anomalies, to conform to the string context, though 
the term ``Wess-Zumino'' (WZ) would probably be more adequate and sufficient for our analysis. The corresponding counterterm will be denoted, GS or WZ, with no distinction.} 
Most of the consequent developments were concentrated around exploiting this idea
in conjunction with supersymmetry and the Froggatt-Nielsen mechanism \cite{FNiel} in order to explain the 
mass hierarchies in the Yukawa sector of the SM \cite{RRR}, supersymmetry breaking \cite{BD},
inflation \cite{Dinfl} and axion physics \cite{GKN}, in all of which the 
presence of the anomalous $U(1)$ is a crucial ingredient. In the context of theories with extra 
dimensions the analysis of anomaly localization and of anomaly inflow has also been at the center of interesting developments \cite{Quiros}, \cite{Hill}. The recent explosion of string model building, in particular in the context
of orientifold constructions and intersecting branes \cite{Orient,genopen}
but also in the context of the heterotic string \cite{Alon},
have enhanced even more the interest in anomalous $U(1)$ models. 
There are a few universal characteristics that these 
vacua seem to possess. One is the presence of $U(1)$ gauge symmetries
that do not appear in the SM \cite{IRU,AKR}. 
In realistic four dimensional heterotic string vacua the SM gauge group 
comes as a subgroup of the ten-dimensional $SO(32)$ or $E_8\times E_8$ 
symmetry \cite{GSW},
and in practice there is at least one anomalous $U(1)$ factor that appears
at low energies, tied to the SM sector in a particular way, which we will summarize
next. For simplicity and reasons of tractability we concentrate on the simplest non-trivial 
case of a model with gauge group $SU(3)_C\times SU(2)_W\times U(1)_Y\times U(1)_B$ 
where $Y$ is hypercharge
and $B$ is the anomalous gauge boson and with the fermion spectrum 
that of the SM. The mass term for the anomalous $U(1)_B$ appears through
a St\"{u}ckelberg coupling \cite{AKR,GIIQ,A1} and the cancellation of its anomalies is due to
four dimensional axionic and Chern-Simons terms (in the open 
string context see the recent works \cite{AKR,AK1,CIK,ABDK}).

Despite of all this theoretical insight both from the top-down and bottom-up approaches,
the question that remains open is how to make concrete contact with experiment.
 However, as mentioned above,
in models with anomalous $U(1)$'s one should quite generally expect the presence
of a physical axion-like field $\chi$ and in fact in any decay that involves a non-vanishing fermion triangle like the 
decay $Z^*, Z^{'*}\longrightarrow \g\g$,  $Z, Z'\longrightarrow Z\g$ etc., one should be able to see
traces of the anomalous structure \cite{CIK,ABDK,ClauNikLet,CIM1}. 
In this paper we will mostly concentrate on the gauge boson decays which, even though
hard to measure,  contain clear differences with respect to the SM - as is the case of
the $Z^*\longrightarrow \g\g$ decay -  and in addition with respect to anomaly free $U(1)$ extensions
- like the $Z^{'*}\longrightarrow \g\g$ decay - for example.
 
In \cite{CIK} a theory which extends the SM with this minimal structure 
(for essentially an arbitrary number of extra $U(1)$ factors)
was called "Minimal Low Scale Orientifold Model" or MLSOM for short,
because in orientifold constructions one typically finds multiple anomalous $U(1)$'s.
Here, even though we discuss the case of a single anomalous $U(1)$ 
which could also originate from heterotic vacua or some field theory extension of the SM, 
we will keep on using the same terminology keeping in mind 
that the results can apply to more general cases.
We finally mention that other similar constructions with emphasis on other phenomenological 
signatures of such models have appeared before in \cite{AK1,AKRT,KN1,KN}. A perturbative study of the renormalization of these types of models is 
in \cite{MM}. Other features of these models, in view of the recent activity 
connected to the claimed PVLAS result \cite{sette}, have been discussed in 
\cite{CIM1}.

Our work is organized as follows. In the first sections we will specialize the analysis of 
\cite{CIK} to the case of an extension of the SM that contains one additional anomalous 
abelian $U(1)$, with an abelian structure of the form $U(1)_Y\times U(1)_B$, that we will analyze 
in depth. We will determine the structure of the entire lagrangean and fix the 
counterterms in the 1-loop anomalous effective action which are necessary 
to restore the gauge invariance of the model at quantum level. 
The analysis that we provide is the generalization of 
what is discussed in \cite{CIM1} that was devoted primarily to the analysis of 
anomalous abelian models and to the perturbative organization of the corresponding effective 
action. 
After determining the axion lagrangian and after discussing Higgs-axion mixing 
in this extension of the SM, we will focus our attention on an analysis of the contributions  
to a simple process $(Z\to \gamma \gamma)$. Our analysis, in this case, aims to provide an 
example of how the new contributions included in the effective action - in the form of one loop counterterms 
that restore unitarity of the effective action - modify the perturbative structure of the process. A detailed phenomenological analysis is beyond the scope of this work, since it requires, to be practically useful for searches at the LHC, a very accurate determination of the QCD and electroweak background 
around the Z/Z' resonance. We hope to return to a complete analysis of 3-linear gauge interactions in this class of models in the near future.

%
\section{Effective models at low energy:  
the $SU(3)_C \times SU(2)_W \times U(1)_Y \times U(1)_B$ case} 
%

We start by briefly recalling the main features of the MLSOM starting from 
the expression of the lagrangean which is given by
\begin{eqnarray}
{\cal L}\; =&-&\frac{1}{2} Tr\;[ F^{G}_{\mu\nu}F^{G\mu\nu} ] - \frac{1}{2} Tr[ \; F^{W}_{\mu\nu} F^{W\mu\nu}]
-\frac{1}{4} F^{B}_{\mu\nu} F^{B\mu\nu} -\frac{1}{4} F^{Y}_{\mu\nu} F^{Y\mu\nu}   \nonumber \\
&+&| ( \partial_{\mu} + i g^{}_{2} \frac{ \tau^j }{ 2 } W_{\mu}^j
+i g^{}_{Y} q^{Y}_{u} A_{\mu}^{Y} +i g^{}_{B} \frac{q^{B}_{u}}{2} B_{\mu} ) H_u|^2    \nonumber\\
&+& | ( \partial_{\mu} + i g^{}_{2} \frac{  \tau^j }{ 2 }  W_{\mu}^j
+i g^{}_{Y} q^{Y}_{d} A_{\mu}^{Y} +i g^{}_{B} \frac{q^{B}_{d}}{2} B_{\mu} ) H_d|^2
\nonumber \\
&+&\overline{Q}_{Li} \, i \gamma^{\mu} \left( \partial^{}_{\mu} 
+i g^{}_{3} \frac{\lambda^{a}}{2} G^{a}_{\mu}+ i g^{}_{2} \frac{\tau^{j}}{2} W^{j}_{\mu} 
+ i g^{}_{Y} q^{(Q_L)}_{Y} A^{Y}_{\mu} + i g^{}_{B} q^{(Q_L)}_{B} B_{\mu} \right) Q_{Li} \nonumber\\
&+& \overline{u}_{Ri}  \, i \gamma^{\mu}  \left( \partial_{\mu} + i g^{}_{Y} q^{(u_R)}_{Y}A^{Y}_{\mu} 
+ i g^{}_{B} q^{(u_R)}_{B}  B_{\mu}   \right) {u}_{Ri} 
+ \overline{d}_{Ri}  \, i \gamma^{\mu}  \left( \partial_{\mu} + i g^{}_{Y} q^{(d_R)}_{Y}A^{Y}_{\mu} 
+ i g^{}_{B} q^{(d_R)}_{B}   B_{\mu}  \right)    {d}_{Ri} \nonumber \\
&+& \overline{L}_{i} \, i \gamma^{\mu} \left( \partial^{}_{\mu} + i g^{}_{2} \frac{\tau^{j}}{2} W^{j}_{\mu} 
+ i g^{}_{Y} q^{(L)}_{Y} A^{Y}_{\mu} + i g^{}_{B} q^{(L)}_{B} B_{\mu} \right) L_{i} \nonumber\\
&+& \overline{e}^{}_{Ri} \, i \gamma^{\mu}  \left( \partial_{\mu} + i g^{}_{Y} q^{(e_R)}_{Y}A^{Y}_{\mu} 
+ i g^{}_{B} q^{(e_R)}_{B} B_\mu\right)  {e}_{Ri} +
\overline{\nu}_{Ri} \, i \gamma^{\mu} \left( \partial_{\mu} + i g^{}_{Y} q^{(\nu_R)}_{Y}A^{Y}_{\mu} 
+ i g^{}_{B} q^{(\nu_R)}_{B}  B_\mu   \right) {\nu}_{Ri}\nonumber \\
&-&  \Gamma^{d} \, \overline{Q}_{L} H_{d} d_{R} - \Gamma^{u} \, \overline{Q}_{L} (i \sigma_2 H^{*}_{u}) u_{R} 
+ c.c. \nonumber\\
&-&   \Gamma^{e} \, \overline{L} H_{d} {e}_{R} - \Gamma^{\nu} \, \overline{L} (i \sigma_2 H^{*}_{u}) \nu_{R} 
+ c.c.\nonumber\\
&+& \frac{1}{2}(\partial_{\mu}b + M^{}_{1} B_{\mu})^2 \nonumber\\ 
&+& \frac{C_{BB}}{M} b  F_{B} \wedge F_{B}  
+ \frac{C_{YY}}{M}  b F_{Y} \wedge F_{Y}   + \frac{C_{YB}}{M}  b F_{Y} \wedge F_{B}   \nonumber\\
&+& \frac{F}{M}  b Tr[F^W \wedge F^W]    +  \frac{D}{M}  b Tr[F^G \wedge F^G]   \nonumber\\ 
&+& d_{1} BY \wedge F_{Y} + d_{2}  YB \wedge F_{B}  + c_{1}  \epsilon^{\mu\nu\rho\sigma} B_{\mu} C^{SU(2)}_{\nu\rho\sigma} 
+ c_{2}  \epsilon^{\mu\nu\rho\sigma} B_{\mu} C^{SU(3)}_{\nu\rho\sigma}    \nonumber\\
&+& V(H_u,H_d,b),
\label{action}
\end{eqnarray}
where we have summed over the $SU(3)$ index $a=1,2,...,8$, over the $SU(2)$ index $j=1,2,3$ and over the fermion index $i=1,2,3$ denoting a given generation. 
We have denoted with $F^{G}_{\mu\nu}$ the field-strength for the
gluons and with $F^{W}_{\mu\nu}$ the field strength of the weak gauge bosons $W_{\mu}$. 
$F^{Y}_{\mu\nu}$ and $F^{B}_{\mu\nu}$ are the field-strengths related to the abelian hypercharge and the extra abelian gauge boson, B, which has anomalous interactions with a typical generation of 
the Standard Model.
The fermions in eq. ($\ref{action}$) are either left-handed or right-handed Dirac spinors $f_L$, 
$f_R$ and they fall in the usual $SU(3)_C$ and 
$SU(2)_W$ representations of the Standard Model.
The additional anomalous $U(1)_{B}$ is accompanied by a shifting St\"{u}ckelberg axion $b$. 
The $c_i$, $i=1,2$, are the coefficients of the Chern-Simons trilinear interactions \cite{CIK,ABDK} and we have also introduced a mass term 
$M_{1} $ at tree level for the B gauge boson, which is 
the St\"{u}ckelberg term. As usual, the hypercharge is anomaly-free and its 
embedding in the so called ``D-brane basis'' has been discussed extensively  
in the previous literature \cite{IRU,AKRT,GIIQ}. Most of the features of the orientifold construction are preserved, 
but we don't work with the more general multiple $U(1)$ structure since our goal 
is to analyze as close as possible this model making contact with direct phenomenological 
applications, although our results and methods can 
be promptly generalized to more complex situations.  

Before moving to the more specific analysis presented in this work, some comments are in order concerning the possible range of validity of effective actions of this type and the relation between the value of the cutoff 
parameter $\Lambda$ and the St\"uckelberg mass $M_1$. This point has been addressed before in 
great detail in \cite{P} and we omit any further elaboration, quoting the result. Lagrangeans 
containing dimension-5 operators in the form of a Wess-Zumino term may have a range of validity constrained 
by  $M_1 \geq g_1 g^2/(64 \pi^3) a_n \Lambda$, where $g_1$ is the coupling at the chiral 
vertex where the anomaly $a_n$ is assigned and g is the coupling constant of the other two 
vector-like currents in a typical AVV diagram. More quantitatively, this bound can be reasonably assumed 
to be of the order of $10^5$ GeV, by a power-counting analysis. Notice 
that the arguments of \cite{P}, though based on the picture of ``partial decoupling'' 
of the fermion spectrum, in which the pseudoscalar field is the phase of a heavier Higgs, 
remain fully valid in this context (see \cite{P} for more details).  
The actual value of $M_1$ is left undetermined, although in the context of string model building there are 
suggestions to relate them to specific properties of the compactified extra dimensions (see for instance \cite{IRU,GIIQ}).

%
%
\section{The effective action of  the MLSOM  with a single anomalous $U(1)$}
\begin{figure}[t]
{\centering \resizebox*{12cm}{!}{\rotatebox{0}
{\includegraphics{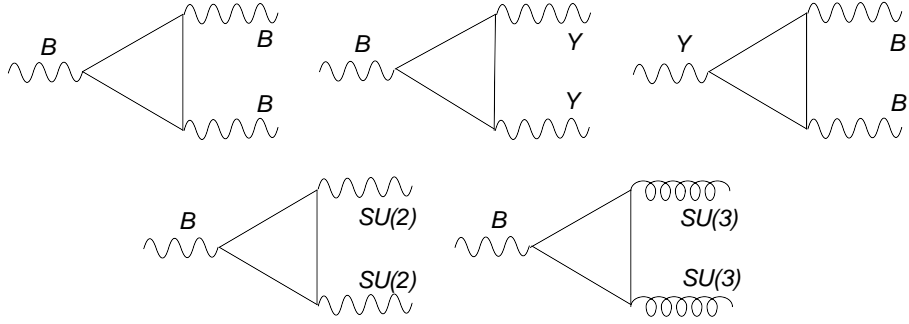}}}\par}
\caption{Anomalous triangle diagrams for the MLSOM.}
\label{anomaly2}
\end{figure}

%
%
%
Having derived the essential components of the classical lagrangean of the model, 
now we try to extend our study to the quantum level, determining the 
anomalous effective action both for the abelian and the non-abelian sectors, fixing 
the $D$, $F$ and $C$ coefficients in front of the Green-Schwarz terms in eq.~\ref{action}. 
Notice that the only anomalous 
contributions to ${\cal S}_{an}$ in the Y-basis before symmetry breaking come from the triangle diagrams depicted in 
Fig.~\ref{anomaly2}.

Since hypercharge is anomaly-free, the only relevant non-abelian 
anomalies to be canceled are those involving one boson $B$ with two $SU(2)_W$ bosons, 
or two $SU(3)_C$ bosons, while the abelian anomalies are those containing three $U(1)$ bosons, 
with the $Y^{3}$ triangle excluded by the hypercharge assignment. These $(BSU(2)SU(2))$ and 
$(BSU(3)SU(3))$ anomalies must be canceled 
respectively by Green-Schwarz terms of the kind
$$F \,b \, Tr[F^W \wedge F^W], \qquad D \,b\, Tr[F^G \wedge F^G],$$
with $F$ and $D$ to be fixed by the conditions of gauge 
invariance. In the abelian sector we have 
to focus on the BBB, BYY and YBB triangles which generate anomalous contributions that need to be 
canceled, respectively, by the Green-Schwarz terms $C_{BB}\,  b \, F^{B} 
\wedge F^{B}$, $C_{YY} \, b \, F^{Y} \wedge F^{Y}$ and $C_{YB} \, b \, F^{Y} \wedge F^{B}$. 
Denoting by ${\cal S}_{YM}$ the anomalous effective action involving 
the classical non-abelian terms plus the non-abelian anomalous diagrams, 
and with ${\mathcal S}_{ab}$ the analogous abelian one, 
the complete anomalous effective action is given by

\beqn
 {\mathcal S}_{eff} &=& {\mathcal S}_0+ {\mathcal S}_{YM} + {\mathcal S}_{ab}   
\eeqn
with ${\mathcal S}_0$ being the classical lagrangean and

\beqa
{\mathcal S}_{YM}&=& \int dx\, dy \, dz 
 \left(\frac{1}{2!} T^{\lambda \mu \nu, ij}_{BWW}(z,x,y) B^{\lambda}(z) W^{\mu}_{i}(x) 
W^{\nu}_{j}(y) + \frac{1 }{2!} T^{\lambda \mu \nu, ab}_{BGG}(z,x,y) B^{\lambda}(z) G^{\mu}_{a}(x) G_{b}^{\nu}(y) \right), \nonumber \\
\eeqa

\beqa
{\mathcal S}_{ab} &=& \int dx\, dy \, dz 
 \left( \frac{1}{ 3! } T^{\lambda \mu \nu}_{BBB}(z,x,y) B^{\lambda}(z) B^{\mu}(x) B^{\nu}(y) 
 + \frac{1}{2!} T^{\lambda \mu \nu}_{BYY}(z,x,y) B^{\lambda}(z) {Y}^{\mu}(x) {Y}^{\nu}(y) \right. 
\nonumber \\
&& \left. \qquad \qquad \qquad + 
 \frac{1}{2!} T^{\lambda \mu \nu}_{YBB}(z,x,y) {Y}^{\lambda}(z) B^{\mu}(x) B^{\nu}(y) \right).
\label{effeaction}
\eeqa
The corresponding 3-point functions, for instance,  are given by 
\beqn
 T^{\lambda \mu \nu,\,ij}_{BWW} B^{\lambda} W^{\mu}_{i} W^{\nu}_{j} 
&=&  \langle 0 | T( J^{\lambda, \, f}_{B} 
J^{\mu, \, f}_{W i} J^{\nu, \, f}_{W j})|0 \rangle  B^{\lambda} W^{\mu}_{i} W^{\nu}_{j}  \nonumber\\  
&\equiv&  \langle 0 | T( J^{\lambda, \, f_L}_{B} 
J^{\mu, \, f_L}_{W i} J^{\nu, \, f_L}_{W j}) |0 \rangle  B^{\lambda} W^{\mu}_{i} W^{\nu}_{j}, 
\eeqn
and similarly for the others. Here we have defined the chiral currents 
\beqn
J^{\lambda, f}_{B} = J^{\lambda, f_R}_{B} + J^{\lambda, f_L}_{B} 
= -  g^{}_{B} q^{\,fR}_{B} \, \overline{\psi}_{f} \gamma^{\lambda} P_{R} \psi_{f} 
-  g^{}_{B} q^{\,fL}_{B} \, \overline{\psi}_{f} \gamma^{\lambda} P_{L} \psi_{f}.
\eeqn
The non-abelian W current being chiral
\beqn
J^{\mu, f}_{Wi} \equiv J^{\mu, f_L}_{Wi}   = -  g^{}_{2} \overline{\psi}_{f} \gamma^{\mu} \tau^{i} P_{L} \psi_{f},
\eeqn
it forces the other currents in the triangle diagram to be of the same chirality, as shown in Fig.~(\ref{all_nonabelian}).

\section{Three gauge boson amplitudes and gauge fixing}
%
\begin{figure}[t]
{\centering \resizebox*{15cm}{!}{\rotatebox{0}
{\includegraphics{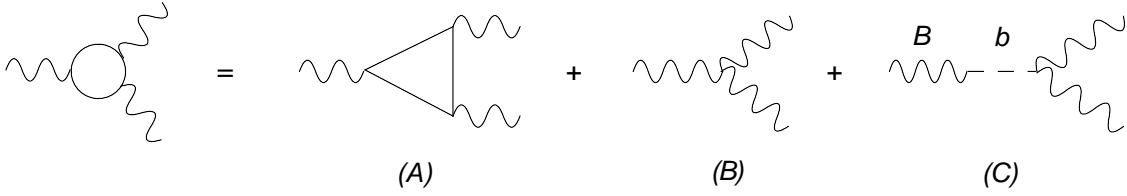}}}\par}
\caption{Contributions to a three abelian gauge boson amplitude before the removal of the 
$B-\partial b$ gauge boson- St\"uckelberg mixing.}
\label{non-abelian}
\end{figure}
%

%
\subsection{The non-abelian sector before symmetry breaking}
Before we get into the discussion of the gauge invariance of the model, it is convenient 
to elaborate on the cancelations of the spurious s-channel poles coming from the gauge-fixing conditions. These are imposed to remove the $\partial b-B$ mixing- in the effective action. We will perform our analysis in the basis of the interaction eigenstates since in this basis recovering gauge 
independence is more 
straightforward, at least before we enforce 
symmetry breaking via the Higgs mechanism. The procedure that we follow 
is to gauge fix the B gauge boson in the symmetric 
phase by removing the $B-\partial b$ mixing (see Fig.~\ref{non-abelian} (C)), so to derive simple Ward identities 
involving only fermionic triangle diagrams and contact trilinear 
interactions with gauge bosons. For this purpose to the St\"uckelberg term
\beq
\frac{1}{2}(\partial_{\mu}b + M^{}_{1} B_{\mu})^2 \nonumber\\ 
\eeq
we add the gauge fixing term  
\beq
\mathcal{L}_{gf}= -\frac{1}{2}\mathcal{G}_B^2 
\eeq
to remove the bilinear mixing, where
\beq
\mathcal{G}_B=\frac{1}{\sqrt{\xi_B}}\left( \partial\cdot B - \xi_B M_1 b\right), 
\eeq
with a propagator for the massive B gauge boson separated in a gauge independent part 
$P_0$ and a gauge dependent one $P_\xi$:
\beqa
\frac{- \,i}{ k^2 - M_1^2} \left( g^{\,\lambda\, \lambda^\prime} - \frac{k^\lambda \, 
k^{\lambda^\prime}}{M_1^2} \right) 
+  \frac{- \,i}{ k^2 - \xi_B \,M_1^2}  \left( \frac{ k^{\lambda} k^{\lambda^\prime}}{M_1^2} 
\right)  
&=& P_0^{\lambda \, \lambda^\prime} +  P_{\xi}^{\lambda \, \lambda^\prime}.
 \eeqa
We will briefly illustrate here 
how the cancelation of the gauge dependence due to $b$ and $B$ exchanges in the 
s-channel goes in this (minimally) gauge-fixed theory. In the exact phase we have no mixing between all 
the $Y,B, W$ gauge bosons and the gauge dependence of the B propagator is canceled by 
the Stueckelberg axion. In the broken phase things get more involved, but essentially the pattern continues to hold. In that case the St\"uckelberg scalar has to be rotated into its physical 
component $\chi $ and the two Goldstones $G_Z$ and $G_{Z'}$ which are linear combinations of $G^0_1$ and 
$G^0_2$. The cancelation of the spurious s-channel poles takes place, in this case, via the combined exchange 
of the $Z$ propagator and of the corresponding Goldstone mode $G_Z $. Naturally 
the GS interaction will be essential for this to happen. 

\begin{figure}[t]
{\centering \resizebox*{10cm}{!}{\rotatebox{0}
{\includegraphics{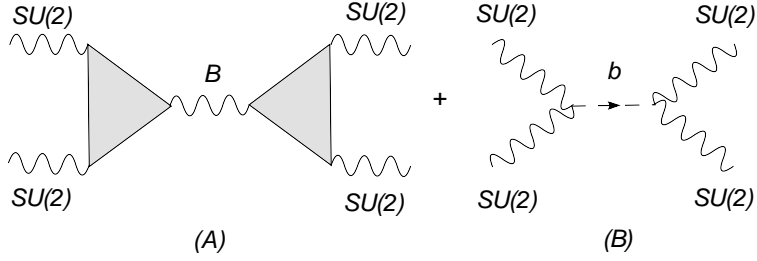}}}\par}
\caption{Unitarity check in SU(2) sector for the MLSOM.}
\label{non-abelian2}
\end{figure}

For the moment we simply work in the exact symmetry phase and in the basis of the interaction eigenstates. 
We gauge fix the action to remove the $B-\partial b$ mixing, but for the rest we set the vev of the scalars 
to zero.
For definiteness let's consider the process $WW \rightarrow WW$ 
mediated by a B boson as shown in Fig.~\ref{non-abelian2}. We denote by a bold-faced 
{\bf V} the $BWW$ vertex, constructed so to have gauge invariance on the W-lines. 
This vertex, as we are going to discuss next, requires a generalized CS 
counterterm to have such a property on the W lines. Gauge invariance on the B line, 
instead, which is clearly necessary 
to remove the gauge dependence in the gauge fixed action, is obtained at a diagrammatical 
level by the the axion exchange (Fig.~\ref{non-abelian2}).
%
%
The expressions of the two diagrams are
\beqn
A^{}_{\xi} + B^{}_{\xi} &=& 
\frac{- i}{k^2 - \xi_B M^2_1} \frac{1}{M^2_1} \Big( k^{\lambda} {\bf V}^{\lambda \mu \nu }_{BWW}(-k^{}_{1}, -k^{}_{2}) \Big) 
\left( k^{\lambda^{\prime}} {\bf V}^{\lambda' \mu' \nu'}_{BWW}( k^{}_{1},  k^{}_{2} ) \right)  
\left(g^{}_{B} g^{\,2}_{2} D^{(L)}_{B} \right)^2  \nonumber\\
&& + \, 4 \times \left( \frac{4 F}{M} \right)^2 
\left(  \frac{i}{ k^2 - \xi_B M^2_1} \right) \varepsilon^{ \mu \nu \alpha \beta} k^{\alpha}_1 k^{\beta}_2  \,
\varepsilon^{\mu' \nu' \alpha' \beta' }  k^{\alpha'}_1 k^{\beta'}_2. 
\eeqn
Using the equations for the anomalies and the correct value for the Green-Schwarz coefficient F given in 
eq.~(\ref{coeff_GS}) (and that we will determine in the next section), we obtain
\beqn
A^{}_{\xi} + B^{}_{\xi} &=& 
\frac{- i}{k^2 - \xi_B M^2_1} \frac{1}{M^2_1} \Big( -  4 a_n \varepsilon k_1 k_2 \Big) 
\Big(  4 a_n \varepsilon' k'_1 k'_2 \Big)  \left( g^{}_{B} g^{2}_{2} D^{(L)}_{B} \right)^2   \nonumber\\
&&+ \frac{64}{M^2}  \frac{M^{2}}{M^{2}_1}
\left(  i g^{}_{B} g^{2}_{2} \frac{a_n}{2} D^{(L)}_{B} \right)^2 
\left(  \frac{i}{ k^2 - \xi_B M^2_1} \right) \varepsilon k_1 k_2 \varepsilon'  k'_1 k'_2 
\eeqn
so that the cancelation is easily satisfied. The treatment of the $SU(3)$ sector is similar and we omit it.

\subsection{The abelian sector before symmetry breaking}
\begin{figure}[t]
{\centering \resizebox*{11cm}{!}{\rotatebox{0}
{\includegraphics{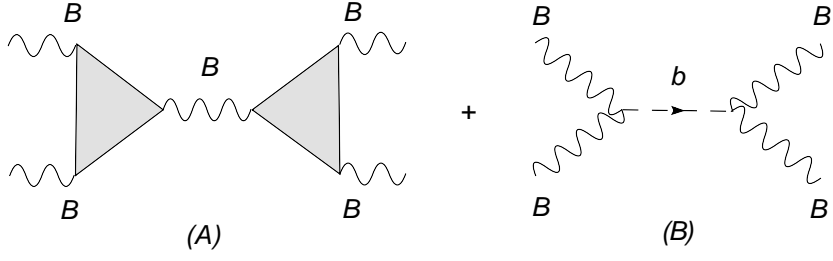}}}\par}
\caption{Unitarity check in abelian sector for the MLSOM.}
\label{BBBanomaly}
\end{figure}

%
%
%
In the abelian sector the procedure is similar. 
For instance, to test the cancelation of the gauge parameter $\xi_{B}$ in a process $BB \rightarrow BB$ mediated by a B gauge boson we sum 
the two gauge dependent contributions coming from the diagrams in Fig.~\ref{BBBanomaly} (we consider only the gauge dependent 
part of the s-channel exchange diagrams)
%
%
\beqn
A^{}_{\xi} + B^{}_{\xi} &=& \frac{- i}{ k^2 - \xi_B M^{2}_{1}} \frac{1}{M^2_1} 
\Big(4 \,  k^{\lambda} { \Delta}^{\lambda \mu \nu}_{\bf AAA}(-k^{}_{1}, -k^{}_{2} ) \Big) 
 \left(4 \, k^{\lambda'} { \Delta}^{\lambda' \mu' \nu'}_{\bf AAA}(k^{}_{1}, k^{}_{2} ) \right) 
\left( g^{3}_{B} D^{}_{BBB} \right)^2    \nonumber\\
&&+ \, 4 \times \left( \frac{4}{M} C_{BB} \right)^{2} \frac{i}{k^2 - \xi_B M^{2}_{1}} 
 \, \varepsilon k_1 k_2  \,   \varepsilon' k'_1 k'_2, 
\eeqn
and cancelation of the gauge dependences implies that the following identity must hold
\beqn
 \frac{16}{M^2_1} \left( \frac{a_n}{3} \right)^2  \left( g^{\,3}_{B} D^{}_{BBB}  \right)^2 
+ 4 \times \left( \frac{4}{M} C_{BB}  \right)^2  = 0,
\eeqn
which can be easily shown to be true after substituting the value of the GS coefficient given in relation (\ref{abelian_BB}). 
%

\begin{figure}[t]
{\centering \resizebox*{11cm}{!}{\rotatebox{0}
{\includegraphics{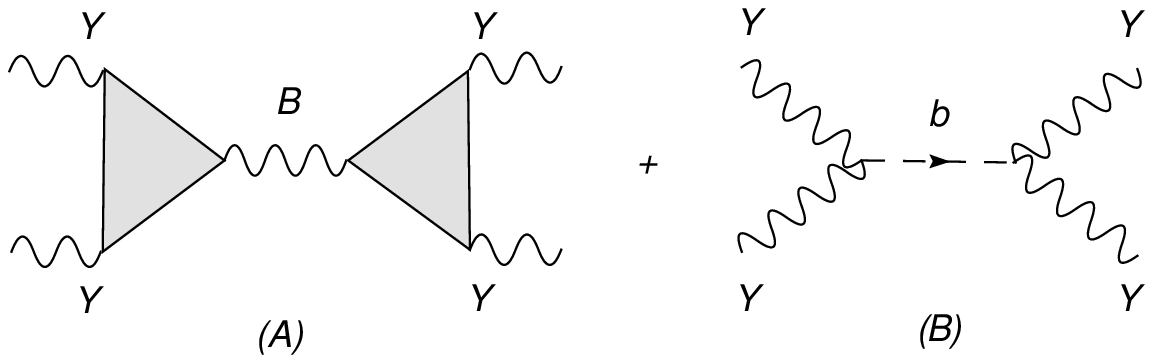}}}\par}
\caption{Unitarity check in abelian sector for the MLSOM.}
\label{BYYanomaly}
\end{figure}

\begin{figure}[t]
{\centering \resizebox*{11cm}{!}{\rotatebox{0}
{\includegraphics{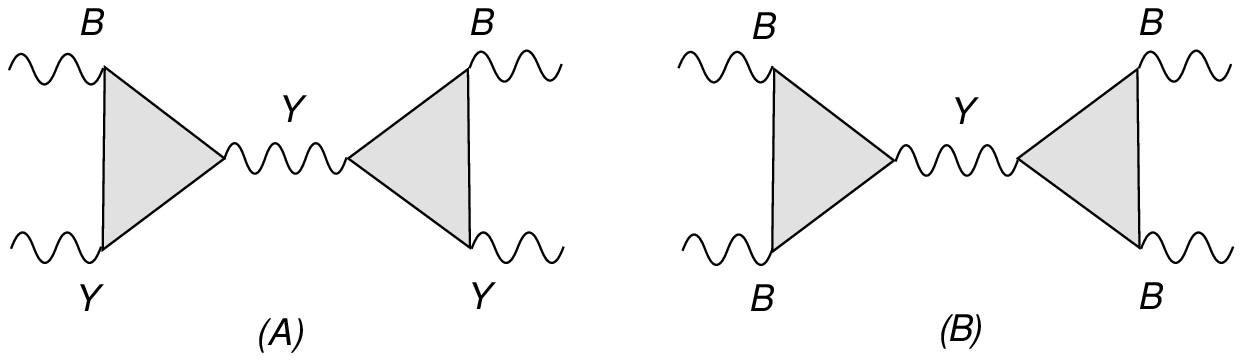}}}\par}
\caption{Unitarity check in abelian sector for the MLSOM.}
\label{Yunitarity}
\end{figure}

%
%
%
%
In Fig.~(\ref{BYYanomaly}) we have depicted the anomalous triangle diagram BYY (A) which 
has to be canceled by the Green-Schwarz 
term $\frac{C_{YY}}{M}bF^{Y} \wedge F^{Y}$, that generates diagram (B). 
In this case the two diagrams give
%

\beqn
A^{}_{\xi} + B^{}_{\xi}  &=&  \frac{- i}{ k^2 - \xi_B M^{2}_{1}} \frac{1}{M^2_1} \Big( k^{\lambda} 
{\bf V}^{\lambda \mu \nu}_{BYY}(-k^{}_{1}, -k^{}_{2} ) \Big) 
 \left( k^{\lambda'} {\bf V}^{\lambda' \mu' \nu'}_{BYY}(k^{}_{1}, k^{}_{2}) \right) \left( g^{}_{B} g^{2}_{Y} D^{}_{BYY} \right)^2
    \nonumber\\
&&+ \, 4 \times \left( \frac{4}{M} C_{YY} \right)^{2} \frac{i}{k^2 - \xi_B M^{2}_{1}} 
 \, \varepsilon k_1 k_2  \,   \varepsilon' k'_1 k'_2. 
\eeqn
The condition of unitarity of the amplitude requires the 
validity of the identity
\beqn
\frac{16}{M^2_1}  \, {a_n}^{2}   \left(  g^{}_{B} g^{\,2}_{Y}   D^{}_{BYY} \right)^2 
+ 4 \times \left( \frac{4}{M} C_{YY}  \right)^2  = 0,
\eeqn
which can be easily checked substituting the value of the GS coefficient $C_{YY}$ given in relation (\ref{abelian_YY}). We will derive the expressions of these
coefficients and the factors of all the other counterterms in the next section. 
The gauge dependences appearing in the diagrams shown 
in Fig.~\ref{Yunitarity} are analyzed in a similar way and we 
omit repeating the previous steps, but it should be obvious by now how the perturbative expansion is organized in terms of tree-level vertices and 1-loop 
counterterms, and how gauge invariance is checked at higher orders when the propagators of the B gauge boson and of the axion b are both present. Notice that in the exact phase the axion $b$ is not coupled to the fermions and the pattern of cancelations to ensure gauge independence, in this specific case, is simplified.   

%
%
%
%
%
%
At this point we pause to make some comments.
The mixed anomalies analyzed above involve a non-anomalous abelian 
gauge boson and the remaining gauge interactions (abelian/non-abelian). 
To be specific, in our model with a single non-anomalous $U(1)$, which is the hypercharge $U(1)_Y$ gauge group, these 
mixed anomalies are those involving triangle diagrams with the $Y$ 
and $B$ generators or the $B$ accompanied by the non-abelian sector.    
Consider, for instance, the $BYY$  triangle, which appears in the $YB \rightarrow YB$ amplitude. There are two options that we can follow. Either we require that the 
corresponding traces of the generators over each generation vanish identically 
\beqn
Tr [ q^{2}_{Y} q_{B} ]   =- 2 \left( -\frac{1}{2}  \right)^{2} q^{(L)}_{B} + (-1)^{2} q^{(e_R)}_{B}  
+ 3 \left[ - 2 \left(\frac{1}{6} \right)^{2}  q^{(Q_L)}_{B} + \left( \frac{2}{3} \right)^{2}    q^{(u_R)}_{B}  
+  \left(  - \frac{1}{3}  \right)^{2} q^{(d_R)}_{B}  \right]   = 0,      \nonumber\\
\eeqn
which can be viewed as a specific condition on the charges of model or, if this is not the case, we require that suitable one-loop counterterms balance the anomalous gauge variation. We are allowed, in other words, to fix the two divergent invariant 
amplitudes of the triangle diagram so that the corresponding Ward identities 
for the $BYY$ vertex and 
similar anomalous vertices are satisfied. This is a condition on the parameterization of the Feynman vertex rather than on the charges and is, in principle, allowed. It is not necessary to have a specific determination of the charges for this to occur, as far as the counterterms are fixed accordingly.
For instance, in the abelian sector the diagrams in question are
\beqn
 YB  \rightarrow YB \,\,\mbox{mediated by Y} \,\, &\propto& \,\,  Tr[ q^2_Y q_B]  \,\,   \nonumber\\
 YY  \rightarrow YY \mbox{mediated by B}\,\, &\propto& \,\,  
Tr[ q^2_Y  q_B ]   \,\,   \nonumber\\
  BB  \rightarrow BB \,\,  \mbox{mediated by Y}    \,\,   &\propto&   \,\, 
Tr[ q_Y q^2_B]  \,\,    \nonumber\\
 YB  \rightarrow YB  \,\, \mbox{mediated by B}  \,\, &\propto&   \,\,
Tr[ q_Y  q^2_B ].    \,\,   \nonumber\\
\eeqn
In the MLSOM these traces are, in general, 
non vanishing and therefore we need to introduce 
defining Ward identities to render the effective action anomaly free.

\section{Ward Identities, Green-Schwarz and Chern-Simons counterterms 
in the St\"uckelberg phase} \label{WI_discuss}
Having discussed the structure of the theory in the basis of 
the interaction eigenstates, we come now to identify the coefficients 
needed to enforce cancelation of the anomalies in the 1-loop effective action. 
In the basis of the physical gauge bosons we will be dropping, with this choice, 
a gauge dependent ( $B\partial b$ mixing) term that is vanishing for 
physical polarizations. At the same time, for exchanges of virtual gauge bosons, the gauge 
dependence of the corresponding propagators is canceled by the associated Goldstone exchanges.  

\begin{figure}[t]
{\centering \resizebox*{16cm}{!}{\rotatebox{0}
{\includegraphics{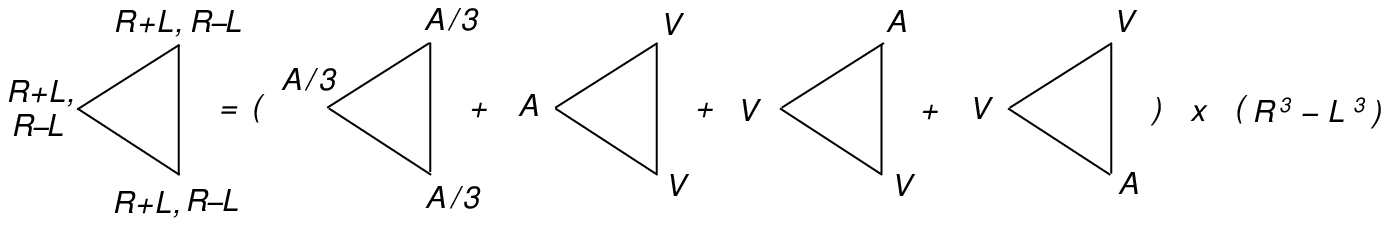}}}\par}
\caption{All the anomalous contributions to a triangle diagram in the abelian sector for 
generic vector-axial vector trilinear interactions in the massless fermion case}
\label{all_abelian}
\end{figure}

Starting from the non abelian contributions, the $BWW$ amplitude, we separate the charge/coupling constant dependence of a 
given diagram from the rest of its 
parametric structure ${\bf T}$ using, in the $SU(2)$ case, the relations 
\beqn
 T^{\lambda \mu \nu,ij}_{BWW}  B^{\lambda} W^{\mu}_{i} W^{\nu}_{j}   &=&g^{}_{B} g^{\,2}_{2} \sum_{i,j} 
Tr[\tau^{}_{i} \tau^{}_{j}] \frac{1}{8} Tr[q^{L}_{B}] 
\,{\bf{T}}^{\,\lambda \mu \nu} B^{\lambda} W^{\mu}_{i} W^{\nu}_{j} \nonumber\\
  &=&  \frac{1}{2}  g^{}_{B} g^{\,2}_{2}  \sum_{i}  
\,D^{(L)}_{B} \,{\bf{T}}^{\,\lambda \mu \nu } B^{\lambda} W^{\mu}_{i} W^{\nu}_{i},   
\label{anom_coeff}
\eeqn
having defined $D^{(L)}_{B} =\frac{1}{8} Tr[q^{L}_{B}] = - \frac{1}{8} \sum_{f} q^{fL}_{B}$ and 
${\bf{T}}^{\lambda \mu \nu}$ is the 3-point function in configuration space, with all the couplings 
and the charges factored out, symmetrized in $\mu \nu$. 
Similarly, for the coupling of $B$ to the gluons we obtain
\beqn
 T^{\lambda \mu \nu,ab}_{BGG} B^{\lambda} G^{\mu}_{a} G^{\nu}_{b} 
&=& g^{}_{B} g^{\,2}_{3} \sum_{a,b} \,Tr[T^{}_{a} T^{}_{b}]  \frac{1}{8} Tr[q^{L}_{B}] 
\,{\bf{T}}^{\,\lambda \mu \nu } B^{\lambda} G^{\mu}_{a} G^{\nu}_{b} \nonumber\\
&=&\frac{1}{2}  g^{}_{B} g^{\,2}_{3}  \sum_{a} D^{(L)}_{B} 
\,{\bf{T}}^{\,\lambda \mu \nu } B^{\lambda} G^{\mu}_{a} G^{\nu}_{a}, 
\eeqn
while the abelian triangle diagrams are given by
\beqn
 T^{\lambda \mu \nu}_{BBB} B^{\lambda} B^{\mu} B^{\nu} 
&=& g^{\,3}_{B} \, \frac{1}{8} Tr[q^{\,3}_{B}] \, {\bf T}^{\,\lambda \mu \nu} B^{\lambda} B^{\mu} B^{\nu}  
= g^{\,3}_{B} \, D_{BBB} \,{\bf{T}}^{\,\lambda \mu \nu} B^{\lambda} B^{\mu} B^{\nu}, \label{BBBvertex}   \\
 {T}^{\lambda \mu \nu}_{BYY} B^{\lambda} Y^{\mu} Y^{\nu} &=& g^{}_{B} g^{\,2}_{Y} \,\frac{1}{8} Tr[q^{}_{B} q^{\,2}_{Y}] \, 
{\bf{T}}^{\,\lambda \mu \nu}  B^{\lambda} Y^{\mu} Y^{\nu}     \nonumber\\
& =& g^{}_{B} g^{\,2}_{Y} \, D_{BYY} \, {\bf{T}}^{\,\lambda \mu \nu}  B^{\lambda} Y^{\mu} Y^{\nu}, \\
T^{\lambda \mu \nu}_{YBB} Y^{\lambda} B^{\mu} B^{\nu} &=& g^{}_{Y} g^{\,2}_{B} \,\frac{1}{8} Tr[q^{}_{Y} q^{\,2}_{B}] \, 
{\bf{T}}^{\,\lambda \mu \nu} Y^{\lambda} B^{\mu} B^{\nu}   \nonumber\\
& =& g^{}_{Y} g^{\,2}_{B} \, D_{YBB} \, {\bf{T}}^{\,\lambda \mu \nu} Y^{\lambda} B^{\mu} B^{\nu},
\eeqn
with the following definitions for the traces (see also the discussion in the Appendix)
\beqn
D^{}_{BBB}&=&   \frac{1}{8} Tr[q^{3}_{B}] = \frac{1}{8} \sum_{f} \left[ (q^{fR}_{B})^{3}
 - (q^{fL}_{B})^{3}  \right],    \\
D^{}_{BYY}&=&    \frac{1}{8} Tr[q^{}_{B} q^{2}_{Y}] =
                \frac{1}{8} \sum_{f} \left[ q^{fR}_{B} (q^{fR}_{Y})^{2}  - q^{fL}_{B} (q^{fL}_{Y})^{2}  \right],       \\
D^{}_{YBB}&=&    \frac{1}{8} Tr[q^{}_{Y} q^{2}_{B}] =
                \frac{1}{8} \sum_{f} \left[ q^{fR}_{Y} (q^{fR}_{B})^{2}  - q^{fL}_{Y} (q^{fL}_{B})^{2}   \right].      
\label{DDD}
\eeqn
The ${\bf T}$ vertex is given by the usual combination of vector and axial-vector components 

\beqn
  {\bf T}^{\lambda \mu \nu} = T^{\lambda \mu \nu}_{\bf AAA} + T^{\lambda \mu \nu}_{\bf AVV} 
+ T^{\lambda \mu \nu}_{\bf VAV} + T^{\lambda \mu \nu}_{\bf VVA},
\eeqn

and we denote by ${\bf \Delta}(k_1,k_2)$ its expression in momentum space 

\beqn
(2\pi)^4 \delta(k-k_1-k_2) {\bf \Delta}^{\lambda \mu \nu}(k_1, k_2) 
= \int dx dy dz e^{ik_1 \cdot x + i k_2 \cdot y - i k \cdot z}  \,
{\bf{T}}^{\lambda \mu \nu}(z,x,y). 
\label{configur}
\eeqn
We denote similarly with ${\bf \Delta}_{\bf AVV}^{\lambda \mu \nu},{\bf \Delta}_{\bf VAV}^{\lambda \mu \nu},{\bf \Delta}_{\bf VVA}^{\lambda \mu \nu}$ the momentum space expressions of the 
corresponding x-space vertices ${\bf T}_{\bf AVV}^{\lambda \mu \nu},{\bf T}_{\bf VVA}^{\lambda \mu \nu}, {\bf T}_{\bf VAV}^{\lambda \mu \nu}$ respectively.
\begin{figure}[t]
{\centering \resizebox*{15cm}{!}{\rotatebox{0}
{\includegraphics{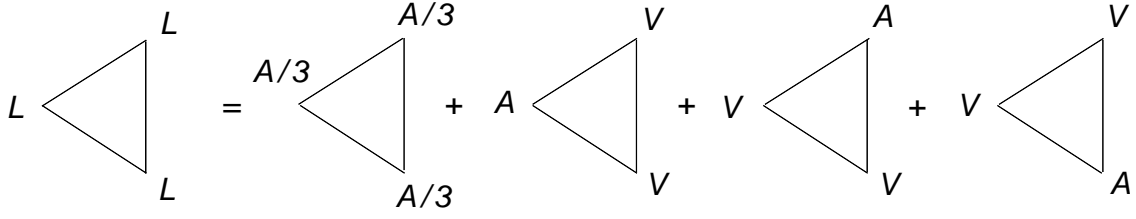}}}\par}
\caption{All the anomalous electroweak contributions to a triangle diagram in the non-abelian sector in the massless fermion case}
\label{all_nonabelian}
\end{figure}
As illustrated in Fig.~(\ref{all_nonabelian}) and Fig.~(\ref{all_abelian}), the complete structure of 
${\bf T }$ is given by 
\beqn
{\bf \Delta}^{\lambda \mu \nu}(k_1, k_2)  &=& \frac{1}{3} \left[ {\Delta}^{\lambda \mu \nu}(- 1/2, k_1, k_2) 
+ { \Delta}^{ \mu \nu \lambda}(-1/2, k_2, - k)  + {\Delta}^{ \nu \lambda \mu}(-1/2, - k, k_1)  \right]   \nonumber\\
&&+ { \Delta}^{\lambda \mu \nu}(-1/2, k_1, k_2) 
+ { \Delta}^{ \mu \nu \lambda }(-1/2, k_2, -  k) + { \Delta}^{ \nu \lambda \mu}(-1/2, - k, k_1)    \nonumber\\
&=& \frac{4}{3} \left[   {\Delta}^{\lambda \mu \nu}(-1/2, k_1, k_2) 
+ { \Delta}^{ \mu \nu \lambda}(-1/2, k_2, - k)  + { \Delta}^{ \nu \lambda \mu}(-1/2, - k, k_1)    \right]   \nonumber\\
&=& 4 \Delta^{\lambda \mu \nu}_{\bf AAA},
\label{aaa}
\eeqn
where we have used the relation between the ${\bf \Delta}_{\bf AAA}$ (bold-faced) vertex and the usual 
$\Delta$ vertex, which is of the form ${\bf AVV}$. Notice that 

\beqa
{\bf \Delta}^{\lambda\mu\nu}_{\bf AVV}&=&{ \Delta}^{\lambda\mu\nu}(-1/2,k_1,k_2), \nonumber \\
{\bf \Delta}^{\mu\nu \lambda}_{\bf AVV}&=&{ \Delta}^{\mu\nu\lambda}(-1/2,k_2,-k), \nonumber \\
{\bf \Delta}^{\nu\lambda\mu}_{\bf AVV}&=&{ \Delta}^{\nu\lambda\mu}(-1/2,-k,k_1),
\eeqa
are the usual vertices with conserved vector current (CVC) on two lines and the anomaly on a single axial vertex. 

The {\bf AAA} vertex is 
constructed by symmetrizing the distribution of the anomaly on each of the 
three chiral currents, which is the content of (\ref{aaa}). The same vertex can be obtained from the basic {\bf AVV} vertex 
by a suitable shift, with $\beta= 1/6$, 
and then repeating the same procedure on the other 
indices and external momenta, with a cyclic permutation. We obtain

\beqa
\Delta^{\lambda \mu \nu}_{\bf AAA}(1/6, k_1, k_2) &=& 
\Delta^{\lambda \mu \nu}(-1/2,k_1, k_2) - \frac{i}{4 \pi^{2}} 
\frac{2}{3} \epsilon^{\lambda \mu \nu \sigma}(k_1 - k_2)_{\sigma} \nonumber \\
\eeqa
and its corresponding anomaly equations are given by
\beqn
 k_{ \lambda } \Delta^{  \lambda \mu \nu}_{\bf AAA}(1/6, k_1, k_2) &=&  \frac{a_n}{3} 
\varepsilon^{\, \mu \nu \alpha \beta} k_{1 \alpha} k_{2 \beta} \nonumber\\
 k_{1 \mu } \Delta^{ \lambda \mu \nu }_{\bf AAA}(1/6, k_1, k_2) &=&  \frac{a_n}{3} 
\varepsilon^{\,\lambda \nu \alpha \beta} k_{1\alpha} k_{2 \beta}\nonumber\\
  k_{2 \nu } \Delta^{  \lambda \mu \nu }_{\bf AAA}(1/6, k_1, k_2) &=& \frac{a_{n}}{3} 
\varepsilon^{\,\lambda \mu \alpha \beta} k_{2\alpha} k_{1 \beta},
\eeqn
typical of a symmetric distribution of the anomaly.

These identities are obtained from the general shift-relation
\beqn
  \Delta^{\lambda \mu \nu}(\beta^\prime , k^{}_{1}, k^{}_{2}) =
 \Delta^{\lambda \mu \nu}(\beta, k^{}_{1}, k^{}_{2} ) 
+   \frac{i}{ 4 \pi^{2} }(\beta - \beta^{\prime} ) 
\epsilon^{\lambda \mu \nu \sigma}(k^{}_{1} - k^{}_{2})^{}_{\sigma}. 
\eeqn 
Vertices with conserved axial currents (CAC) can be related to the symmetric {\bf AAA} vertex in a similar way 
\beqn
\Delta^{ \lambda \mu \nu }_{\bf AAA}( + 1/6 , k^{}_{1}, k^{}_{2}) =
 \{ \Delta^{\lambda \mu \nu}(+ 1/2, k^{}_{1}, k^{}_{2} ) \}_{CAC} 
+ \frac{i}{ 4 \pi^{2} } \, \frac{1}{3} \,
\epsilon^{\lambda \mu \nu \sigma}(k^{}_{1} - k^{}_{2})^{}_{\sigma}.
\eeqn

At this point we are ready to introduce the complete vertices 
for this model, which are given 
by the amplitude ($\ref{configur}$) with the addition of 
the corresponding Chern-Simons 
counterterms, were required. These will be determined later in this section 
by imposing the conservation of the 
$SU(2)$, $SU(3)$ and $Y$ gauge currents. Following this definition for all the 
anomalous vertices, 
the amplitudes can then be written as 

\beqn
{\bf{\mathcal V}}^{\lambda \mu \nu, \, aa}_{BGG} B^{\lambda} G^{\mu}_{a} G^{\nu}_{a} &=& 
\frac{1}{2} g^{}_{B} g^{\,2}_{3} D^{(L)}_{B} {\bf T}^{\lambda \mu \nu} B^{\lambda} G^{\mu}_{a} G^{\nu}_{a}
+ c^{}_{2} \epsilon^{ \mu \nu \rho \sigma} B^{}_{\mu} C^{SU(3)}_{\nu \rho \sigma}   \nonumber\\
{\bf {\mathcal V}}^{\lambda \mu \nu, \, ii}_{BWW} B^{\lambda} W^{\mu}_{i} W^{\nu}_{i} &=& 
\frac{1}{2} g^{}_{B} g^{\,2}_{2} D^{(L)}_{B} {\bf T}^{\lambda \mu \nu}  B^{\lambda} W^{\mu}_{i} W^{\nu}_{i}
+ c^{}_{1} \epsilon^{ \mu \nu \rho \sigma} B^{}_{\mu} C^{SU(2)}_{\nu \rho \sigma}   \nonumber\\
{\bf {\mathcal V}}^{\lambda \mu \nu}_{BYY} B^{\lambda} Y^{\mu} Y^{\nu} &=& 
 g^{}_{B} g^{\,2}_{Y} D^{}_{BYY} {\bf T}^{\lambda \mu \nu} B^{\lambda} Y^{\mu} Y^{\nu}
+ d^{}_{1} B Y \wedge F^{}_{Y}  \nonumber\\
{\bf{\mathcal V}}^{\lambda \mu \nu}_{YBB} Y^{\lambda} B^{\mu} B^{\nu}  &=&  
g^{}_{Y} g^{\,2}_{B} D^{}_{YBB} {\bf T}^{\lambda \mu \nu}  Y^{\lambda} B^{\mu} B^{\nu} 
+ d^{}_{2} Y B \wedge F^{}_{B},  \nonumber\\
{\bf{\mathcal V}}^{\lambda \mu \nu}_{BBB} B^{\lambda} B^{\mu} B^{\nu}  &=&  
g^{\,3}_{B} D^{}_{BBB} {\bf T}^{\lambda \mu \nu}  B^{\lambda} B^{\mu} B^{\nu} \nonumber\\
\label{definingv}
\eeqn
which are the anomalous vertices of the effective action, corrected when necessary by 
suitable CS interactions in order to conserve all the gauge currents at 1-loop.

Before we proceed with our analysis, which has the goal to determine explicitly the 
counterterms in each of these vertices, we pause for some practical considerations. It is clear 
that the scheme that we have followed in order 
to determine the structure of the vertices of the effective action has been 
to assign the anomaly only to the chiral vertices and to impose conservation of the vector 
current. There are regularization schemes in the literature that enforce this principle, the most famous one 
being dimensional regularization with the t'Hooft Veltman prescription for $\gamma_5$ 
(see also the discussion in part 1).  In this scheme the anomaly is equally distributed for 
vertices of the form {\bf AAA} and is assigned only to the axial-vector vertex in triangles of the form {\bf AVV} and similar. Diagrams of the form {\bf AAV } are zero by Furry's theorem, being equivalent to {\bf VVV}. 

We could also have proceeded in a different way, for instance 
by defining each $\mathcal{V}$, for instance ${\mathcal V}_{BYY}$, to have an anomaly only 
on the B vertex and not on the Y vertices, even if Y has both a vector and an axial-vector 
components at tree level and is, indeed, a chiral current. This implies 
that at 1-loop the chiral projector has to be moved from the Y to to the B vertex ``by hand'', no matter if it  appears on the Y current or on the B current, rendering the Y current effectively vector-like at 
1 loop. This is also what a CS term does. In both cases we are anyhow bond 
to define separately the 1-loop vertices as new entities, unrelated to 
the tree level currents. However, having explicit Chern-Simons counterterms renders the treatment compatible with dimensional regularization in the t'Hooft-Veltman prescription. 
It is clear, however, that one way or the other, the quantum action is not fixed at classical level since the counterterms are related to quantum effects 
and the corresponding Ward identities, which force 
the cancelation of the anomaly to take place in a completely new way respect to the SM case, 
are indeed {\em defining conditions} on the theory.  

Having clarified this subtle point, we return to the determination of the gauge invariance 
conditions for our anomalous vertices. 

Under B-gauge transformations we have the following variations (singlet anomalies) of the effective action 
\beqa
\frac{1}{2!} \delta_B < T_{ BWW} B W W> =  i \frac{ g^{}_{B} g^{\,2}_{2} }{2!}  \, \frac{4}{3}a_n \, \frac{1}{4} \langle \theta_B
 F^{W}_{i} \wedge F^{W}_{j}  \rangle \, Tr[ \tau^{i}  \tau^{j} ] \,  D^{(L)}_{B},   
\label{MLSOM_2}
\eeqn
\beqn
\frac{1}{2!} \delta_B < T_{ BGG } B G G> =  i \frac{g^{}_{B} g^{\,2}_{3} }{ 2!} \,   \frac{4}{3}a_n  \, \frac{1}{4} \langle   \theta_B
  F^{G}_{a} \wedge  F^{G}_{b}  \rangle \,  Tr[ T^{a} T^{b} ]  \, D^{(L)}_{B}, 
\label{MLSOM_3}
\eeqa
and with the normalization given by 

\beqa
Tr [ \tau^{i} \tau^{j} ] = \frac{1}{2} \delta^{ij}    \qquad  Tr[T^{a}T^{b}] = \frac{1}{2} \delta^{ab}
\eeqa
we obtain 
\beqa
 \frac{1}{2!} \delta_B < T_{B SU(2) SU(2)} B W W > &=&  i \frac{g^{}_{B} g^{\,2}_{2}}{2!} \, \frac{a_{n}}{6}  \, 
\langle \theta_B  F^W_i \wedge F^W_i   \rangle   D^{(L)}_{B},  \\
\frac{1}{2!} \delta_B < T_{B SU(3) SU(3)} B G G>  &=&  i \frac{g^{}_{B} g^{\,2}_{3}}{2!} \,  \frac{a_{n}}{6}    
 \,  \langle  \theta_B F^G_a \wedge F^G_a \rangle D^{(L)}_{B}.
\eeqa
Note, in particular, that the covariantization of the anomalous contributions requires the entire non-abelian field strengths 
$F^W_{i,\,\mu \nu}$ and $F^G_{a,\,\mu \nu}$
\beqn
F^W_{i, \,\mu \nu} &=& \partial_{\mu} W^{i}_{\nu} - \partial_{\nu} W^{i}_{\mu}  
-  g^{}_{2} \varepsilon_{ijk} W^{j}_{\mu} W^{k}_{\nu} 
=  \hat{F}^{W}_{i,\, \mu \nu}-  g^{}_{2} \varepsilon_{ijk} W^{j}_{\mu} W^{k}_{\nu} \\
F^G_{a,\,\mu \nu} &=& \partial_{\mu} G^{a}_{\nu} - \partial_{\nu} G^{a}_{\mu} 
 -  g^{}_{3} f_{abc} G^{b}_{\mu} G^{c}_{\nu} = \hat{F}^{G}_{a,\, \mu \nu} -  g^{}_{3} f_{abc} G^{b}_{\mu} G^{c}_{\nu}. 
\eeqn
The covariantization of the right-hand-side (rhs) of the anomaly equations takes place 
via higher order corrections, involving correlators with more external gauge lines. It is well known, though, that the cancelation of the anomalies in these higher order 
non-abelian diagrams (in d=4) is only related to the triangle diagram 
(see \cite{CIM1}). 

Under the non-abelian gauge transformations we have the following variations
\beqn
\frac{1}{2!} \delta_{SU(2)} \langle T_{BWW} BWW  \rangle &=& i  \frac{g^{}_{B} g^{\,2}_{2}}{2!} \frac{8}{3}
a^{}_{n} \frac{1}{4} \langle F^{B} \wedge Tr[ \theta   \hat{F}^{W}  ]  \rangle  D^{(L)}_{B}   \\
\frac{1}{2!} \delta_{SU(3)} \langle T_{BGG} BGG  \rangle &=& i  \frac{g^{}_{B} g^{\,2}_{3}}{ 2! } \frac{8}{3}
a^{}_{n} \frac{1}{4} \langle  F^{B} \wedge Tr[ \vartheta \hat{F}^{G} ]  \rangle   D^{(L)}_{B},
\eeqn
where the ``hat'' field strengths $\hat{F}^{W}$ and $\hat{F}^{G}$ refer to the abelian part of the non-abelian field strengths W and G. Introducing the notation
\beqn
Tr[\theta \hat{F}^{W}] &=&  Tr[\tau^{i} \tau^{j}] \theta^{}_{i} \hat{F}^{W}_{j}
 =  \frac{1}{2} \theta^{}_{i} \hat{F}^{W}_{i}   \;\;\,\,\,\,\,\,\,\, i,j = 1,2,3  \\
   Tr[\vartheta \hat{F}^{G}] &=& Tr[T^{a} T^{b}] \vartheta_{a} \hat{F}^{G}_{b} 
=  \frac{1}{2} \vartheta_{a} \hat{F}^{G}_{a}    \,\,\;\;\,\,\,\,\,\,  a,b = 1,2,..,8
\eeqn
the expressions of the variations become
\beqn
\frac{1}{2!} \delta_{SU(2)} \langle T_{BWW} BWW  \rangle &=& i  \frac{g^{}_{B} g^{\,2}_{2}}{2!} 
\frac{a^{}_{n}}{3}  \langle \theta^{}_{i} F^{B} \wedge  \hat{F}^{W}_{i}   \rangle  D^{(L)}_{B}    \\
\frac{1}{2!} \delta_{SU(3)} \langle T_{BGG} BGG  \rangle &=& i  \frac{g^{}_{B} g^{\,2}_{3}}{ 2! }
\frac{a^{}_{n}}{3} \langle \vartheta_{a}  F^{B} \wedge  \hat{F}^{G}_{a}   \rangle   D^{(L)}_{B}.
\eeqn
We have now to introduce the Chern-Simons counterterms for the non-abelian gauge variations 

\beqn
{\mathcal S}^{CS}_{non-ab} = {\mathcal S}^{CS}_{BWW} + {\mathcal S}^{CS}_{BGG} 
= c^{}_{1} \langle \varepsilon^{\mu \nu \rho \sigma} B^{}_{\mu} 
C^{SU(2)}_{\nu \rho \sigma}   \rangle + c^{}_{2} \langle   \varepsilon^{\mu \nu \rho \sigma} B^{}_{\mu} 
C^{SU(3)}_{\nu \rho \sigma}    \rangle,
\eeqn
with the non-abelian CS forms given by
\beqn
C^{SU(2)}_{\mu \nu \rho} &=&  \frac{1}{6} \left[ W^{i}_{\mu} \left( F^W_{i,\,\nu \rho} + \frac{1  }{3} \, g^{}_{2}  
\, \varepsilon^{ijk} W^{j}_{\nu} W^{k}_{\rho}  \right) + cyclic   \right]              ,    \\
C^{SU(3)}_{\mu \nu \rho} &=&  \frac{1}{6} \left[ G^{a}_{\mu} \left( F^G_{a,\,\nu \rho} + \frac{1 }{3} \, g^{}_{3}
\, f^{abc} G^{b}_{\nu} G^{c}_{\rho}  \right) + cyclic  \right],      
\eeqn
whose variations under non-abelian gauge transformations are 
\beqn
\delta_{SU(2)} C^{SU(2)}_{\mu \nu \rho} &=& \frac{1}{6}  \left[ \partial^{}_{\mu} \theta^{i} \,( \hat{F}^{W}_{i, \,\nu \rho}) 
+ cyclic \right],      \\
\delta_{SU(3)} C^{SU(3)}_{\mu \nu \rho} &=& \frac{1}{6} \left[ \partial^{}_{\mu} \vartheta^{a} \,( \hat{F}^{G}_{a,\,\nu \rho}) 
+ cyclic \right].      
\eeqn
The variations of the Chern-Simons counterterms then become 
\beqn
\delta_{SU(2)}  {\mathcal S}^{CS}_{BWW}  &=&  \frac{c^{}_{1}}{2} \, \frac{1}{2} 
\langle \theta^{i} F^{B} \wedge \hat{F}^{W}_{i}  \rangle    \\ 
\delta_{SU(3)}  {\mathcal S}^{CS}_{BGG}   &=& \frac{c^{}_{2}}{2} \,  \frac{1}{2}
\langle \vartheta^{a} F^{B} \wedge \hat{F}^{G}_{a}  \rangle,  
\eeqn
and we can choose the coefficients in front of the CS counterterms 
to obtain anomaly cancelations for the non-abelian contributions

\beqn
c^{}_{1} =  - i g^{}_{B} g^{\,2}_{2}\frac{2}{3} a_n D^{(L)}_{B}  \qquad  
c^{}_{2} =  - i g^{}_{B} g^{\,2}_{3}\frac{2}{3} a_n D^{(L)}_{B}. 
\label{fixx}
\eeqn 
The variations under B-gauge transformations for the related CS counterterms are then given by 
\beqn
\delta^{}_{B} {\mathcal S}^{CS}_{BWW} &=& - \frac{c^{}_{1}}{2} \, \frac{1}{2} \langle \theta^{}_{B}   F^W_i \wedge F^W_i  \rangle    \\
\delta^{}_{B} {\mathcal S}^{CS}_{BGG} &=& - \frac{c^{}_{2}}{2} \, \frac{1}{2} \langle \theta^{}_{B} F^G_{a} \wedge F^G_{a} \rangle,
\eeqn
where the coefficients $c_i$ are given in (\ref{fixx}). The variations under the B-gauge transformations for the $SU(2)$ and $SU(3)$ Green-Schwarz 
counterterms are respectively given by
\beqn
 \frac{F}{M} \, \delta_{B} \langle \, b \, Tr[F^W \wedge F^W]  \, \rangle &=& - F \frac{M_1}{M} \frac{1}{2}  \langle \theta_B 
 F^W_{i} \wedge F^W_{i} \rangle, \\
 \frac{D}{M} \, \delta_{B} \langle \, b \, Tr[ F^G \wedge F^G] \, \rangle &=& - D \frac{M_1}{M} \frac{1}{2}  \langle  \theta_B 
 F^G_{a} \wedge F^G_{a}  \rangle,  
\eeqn
and the cancelation of the anomalous contributions coming from the B-gauge transformations 
determines $F$ and $D$ as
\beqn
F =  \frac{M}{M_1} i   g^{}_{B} g^{\,2}_{2} \, \frac{a_n}{2} \, D^{(L)}_{B},    
\qquad D =  \frac{M}{M_1} i  g^{}_{B} g^{\,2}_{3} \,  \frac{a_n}{2} \,   D^{(L)}_{B}.    
\label{coeff_GS}
\eeqn
%
There are some comments to be made concerning the generalized CS terms responsible for the cancelation of the mixed anomalies. These terms, in momentum space, generate standard trilinear CS interactions, whose momentum structure is exactly the same as that due to the abelian ones (see the appendix of part 1 for more details), plus 
additional quadrilinear (contact) gauge interactions. These will be neglected in our analysis since we will 
be focusing in the next sections on the characterization of neutral tri-linear interactions. 
In processes such as $Z\to \gamma\gamma\gamma$ they re-distribute the anomaly appropriately 
in higher point functions.

For the abelian part ${\mathcal S}_{ab}$ of the effective action we first focus on gauge 
variations on B, obtaining 
\beqn
\frac{1}{3!} \delta_{B}  \langle T^{\lambda \mu \nu}_{ BBB } 
B^{\lambda}(z) B^{\mu}(x) B^{\nu}(y)  \rangle  =  i  \frac{g^{\,3}_{B}}{3!} \frac{4}{3} a_n \frac{3}{4}
\langle F^{B} \wedge F^{B} \theta_B \rangle  \, D^{}_{BBB},   
\eeqn 
\beqn
&&\frac{1}{2!} \delta_{B}  \langle T^{\lambda \mu \nu}_{ BYY  } 
B^{\lambda}(z) Y^{\mu}(x) Y^{\nu}(y)  \rangle =  i  \frac{g^{}_{B} g^{2}_{Y}}{2!} \frac{4}{3} a_n  \frac{1}{4}
\langle F^{Y} \wedge F^{Y} \theta_B \rangle  \,  D^{}_{BYY}, 
\eeqn 
\beqn
&&\frac{1}{2!} \delta_{B}  \langle T^{\lambda \mu \nu}_{ YBB  } 
Y^{\lambda}(z) B^{\mu}(x) B^{\nu}(y)  \rangle =  i  \frac{g^{}_{Y} g^{2}_{B}}{2!} \frac{4}{3} a_{n} \frac{2}{4} 
\langle F^{Y} \wedge F^{B} \theta_B \rangle  \,  D^{}_{YBB}, 
\eeqn 
and variations for $Y$ that give 
\beqn
&&\frac{1}{2!} \delta_{Y}  \langle T^{\lambda \mu \nu}_{ BYY } 
B^{\lambda}(z) Y^{\mu}(x) Y^{\nu}(y)  \rangle =  i  \frac{g^{}_{B} g^{2}_{Y}}{2!} \frac{4}{3} a_{n} \frac{2}{4}
\langle F^{Y} \wedge F^{B} \theta_Y \rangle  \,  D^{}_{BYY}, 
\label{var_Y2}
\eeqn 
\beqn
&&\frac{1}{2!} \delta_{Y}  \langle T^{\lambda \mu \nu}_{ YBB  } 
Y^{\lambda}(z) B^{\mu}(x) B^{\nu}(y)  \rangle =  i  \frac{g^{}_{Y} g^{2}_{B}}{2!}  \frac{4}{3} a_n  \frac{1}{4}
\langle F^{B} \wedge F^{B} \theta_Y \rangle  \,  D^{}_{YBB}. 
\label{var_Y3}
\eeqn 
Also in this case we introduce the corresponding abelian Chern-Simons counterterms 
\beqn
{\mathcal S}^{CS}_{ab} = {\mathcal S}^{CS}_{BYY} + {\mathcal S}^{CS}_{YBB}  
=  d^{}_{1} \langle BY \wedge F^{}_{Y} \rangle +  d^{}_{2} \langle YB \wedge F^{}_{B} \rangle
\eeqn
whose variations are given by
\beqn
\delta_{Y}  {\mathcal S}^{CS}_{BYY}   &=&    \frac{ d^{}_{1} }{2} \langle \theta^{}_{Y} F^{B} \wedge F^{Y} \rangle,   \\
\delta_{Y}  {\mathcal S}^{CS}_{YBB}  &=& -  \frac{d^{}_{2} }{2} \langle \theta^{}_{Y} F^{B} \wedge F^{B}  \rangle,   
\eeqn
and we can fix their coefficients so to obtain the cancelation  of the Y-anomaly 
\beqn
d_{1} = - i g_{B} g^{2}_{Y}\frac{2}{3} a^{}_{n} D^{}_{BYY}  \qquad 
 d_{2} =  i g_{Y} g^{2}_{B}  \frac{ a^{}_{n} }{3}  D^{}_{YBB}.
\label{coefficiente_CS_abeliano}
\eeqn
Similarly, the gauge variation of B in the corresponding Green-Schwarz terms gives
\beqn
  \frac{C_{BB}}{M}  \,   \delta_{B} \langle \, b \, F^{B} \wedge F^{B}  \rangle &=& - C_{BB} \frac{M_1}{M} \langle \theta_{B} 
F^{B}\wedge F^{B} \rangle      \\
 \frac{C_{YY}}{M} \, \delta_{B} \langle \, b \, F^{Y} \wedge F^{Y}  \rangle &=& - C_{YY} \frac{M_1}{M} \langle \theta_{B} 
F^{Y}\wedge F^{Y} \rangle      \\
 \frac{C_{YB}}{M} \,  \delta_{B} \langle \,  b \, F^{Y} \wedge F^{B}  \rangle &=& - C_{YB} \frac{M_1}{M} \langle \theta_{B} 
F^{Y}\wedge F^{B} \rangle 
\eeqn
and on the other hand the B-variations of the fixed CS counterterms are 
\beqn
\delta_{B}     {\mathcal S}^{CS}_{BYY}   &=& - \frac{ d^{}_{1} }{2} \langle \theta^{}_{B} F^{Y} \wedge F^{Y} \rangle,   \\
\delta_{B}   {\mathcal S}^{CS}_{YBB}   &=&   \frac{d^{}_{2} }{2} \langle \theta^{}_{B} F^{Y} \wedge F^{B}  \rangle.
\eeqn
Finally the cancelation of the anomalous contributions from the abelian part of the effective action requires
following conditions 
\beqn
C_{BB} &=&  \frac{M}{M_1} \frac{ i g^{\,3}_{B} }{3!} a_n  D^{}_{BBB}, \label{abelian_BB}  \\
C_{YY} &=&  \frac{M}{M_1} i g^{}_{B} g^{\,2}_{Y} \frac{a^{}_{n}}{2} D^{}_{BYY}, \label{abelian_YY} \\
C_{YB} &=&  \frac{M}{M_1} i g^{}_{Y} g^{\,2}_{B}  \frac{a_{n}}{2} D^{}_{YBB}. \label{abelian_YB} 
\eeqn
Regarding the Y-variations $\propto Tr[q_B q^{2}_{Y}]$ and $\propto Tr[q^{2}_{B} q_{Y}]$, in general these traces are 
not identically vanishing and we introduce the CS and GS counterterms to cancel them. 
Having determined the factors in front of all the counterterms, 
we can summarize the structure of the one-loop anomalous 
effective action plus the counterterms as follows 

\beqn
{\mathcal S} &=&   {\mathcal S}_0 +{\mathcal S}_{an} + {\mathcal S}_{GS} + {\mathcal S}_{CS}  \nonumber\\
&=& {\mathcal S}_0 + \frac{1}{2!} \langle T_{BWW} BWW \rangle +  \frac{1}{2!} \langle T_{BGG} BGG \rangle
 + \frac{1}{3!} \langle T_{BBB} BBB \rangle     \nonumber\\
&&+ \frac{1}{2!} \langle T_{BYY} BYY \rangle + \frac{1}{2!} \langle T_{YBB} YBB \rangle  \nonumber\\
&&+ \frac{C_{BB}}{M} \langle b  F_{B} \wedge F_{B}  \rangle 
+ \frac{C_{YY}}{M} \langle b F_{Y} \wedge F_{Y}  \rangle + \frac{C_{YB}}{M} \langle b F_{Y} \wedge F_{B}  \rangle \nonumber\\
&&+ \frac{F}{M} \langle b Tr[F^W \wedge F^W]  \rangle  +  \frac{D}{M} \langle b Tr[F^G \wedge F^G] \rangle  \nonumber\\ 
&&+ d_{1} \langle BY \wedge F_{Y} \rangle + d_{2} \langle YB \wedge F_{B} \rangle  \nonumber\\
&&+ c_{1} \langle \epsilon^{\mu\nu\rho\sigma} B_{\mu} C^{SU(2)}_{\nu\rho\sigma} \rangle
+ c_{2} \langle \epsilon^{\mu\nu\rho\sigma} B_{\mu} C^{SU(3)}_{\nu\rho\sigma} \rangle,   \nonumber\\
\eeqn
where ${\mathcal S}_0$ is the classical action. At this point we are ready to 
define the expressions in momentum space of the vertices introduced in 
eq. (\ref{definingv}), denoted by {\bf{V}}, obtaining

\beqn
{\bf V}^{\lambda \mu \nu}_{BGG} &=& 4 \, \frac{1}{2}  D^{(L)}_{B} \, g^{}_{B} g^{\,2}_{3}
\, \Delta^{\lambda \mu \nu}_{\bf AAA} ( + 1/6 , k^{}_{1}, k^{}_{2} )  
+  D^{(L)}_{B} \, g^{}_{B} g^{\,2}_{3} \frac{1}{2} \frac{i}{ \pi^{2} } \, \frac{2}{3} \,
\epsilon^{\lambda \mu \nu \sigma}(k^{}_{1} - k^{}_{2})^{}_{\sigma} 
\label{v1}    \\
{\bf V}^{\lambda \mu \nu}_{BWW} &=&4  \, \frac{1}{2} D^{(L)}_{B} \, g^{}_{B} g^{\,2}_{2}
\, \Delta^{\lambda \mu \nu}_{\bf AAA} ( + 1/6 , k^{}_{1}, k^{}_{2} )  
+  D^{(L)}_{B} \, g^{}_{B} g^{\,2}_{2} \frac{1}{2} \frac{i}{ \pi^{2} } \, \frac{2}{3} \,
\epsilon^{\lambda \mu \nu \sigma}(k^{}_{1} - k^{}_{2})^{}_{\sigma} 
\label{v1}    \\
{\bf V}^{\lambda \mu \nu}_{BYY} &=&4  D^{}_{BYY} \, g^{}_{B} g^{\,2}_{Y}
\, \Delta^{\lambda \mu \nu}_{\bf AAA} ( + 1/6 , k^{}_{1}, k^{}_{2} )  
+  D^{}_{BYY} \, g^{}_{B} g^{\,2}_{Y} \frac{i}{ \pi^{2} } \, \frac{2}{3} \,
\epsilon^{\lambda \mu \nu \sigma}(k^{}_{1} - k^{}_{2})^{}_{\sigma} 
\label{v1}    \\
{\bf V}^{\lambda \mu \nu}_{YBB} &=& 4 D^{}_{YBB} \, g^{}_{Y} g^{\,2}_{B}
 \, \Delta^{\lambda \mu \nu}_{\bf AAA} ( + 1/6 , k^{}_{1}, k^{}_{2} )  
-  D^{}_{YBB} \, g^{}_{Y} g^{\,2}_{B}  \frac{i}{  \pi^{2} } \, \frac{1}{3} \,
\epsilon^{\lambda \mu \nu \sigma}(k^{}_{1} - k^{}_{2})^{}_{\sigma}.
\label{v2} \\
{\bf V}^{\lambda \mu \nu}_{BBB} &=& 4 D^{}_{BBB} \, g^{\,3}_{B}
 \, \Delta^{\lambda \mu \nu}_{\bf AAA} ( + 1/6 , k^{}_{1}, k^{}_{2} ).
\label{v3}
\eeqn

where for the generalized CS terms we consider only the trilinear CS interactions whose momentum structure is the same as the abelian ones as already discussed in section \ref{WI_discuss}. The factor 1/2 overall in the non abelian vertices comes from the trace over the generators.
 These vertices satisfy standard Ward identities on 
the external Standard Model lines, with an anomalous Ward identity only on the $B$ line 
\beqn
k^{}_{1\mu} {\bf V}^{\lambda \mu \nu}_{BYY}( k^{}_{1}, k^{}_{2}) &=& 0   \\
k^{}_{2 \nu } {\bf V} ^{\lambda \mu \nu }_{BYY} (  k^{}_{1}, k^{}_{2}) &=& 0   \\
k^{}_{\lambda } {\bf V}^{ \lambda \mu \nu }_{BYY} ( k^{}_{1}, k^{}_{2}) &=& 4
 D^{}_{BYY} g^{}_{B} g^{\,2}_{Y} \, a^{}_{n} \,  \epsilon^{\mu \nu \alpha \beta} k_{1 \alpha} k_{2 \beta},
\eeqn
and obviously the B-currents contain the total anomaly $a^{}_{n}= - \frac{i}{2 \pi^{2}}$. The same anomaly equations given above for 
${\bf V}^{\lambda \mu \nu}_{BYY}$ hold for the ${\bf V}^{\lambda \mu \nu}_{BGG}$ and ${\bf V}^{\lambda \mu \nu}_{BWW}$ vertices but with a 1/2 factor overall. The anomaly equations for the YBB vertex are 
\beqn
k^{}_{1\mu} {\bf V}^{\lambda \mu \nu}_{YBB}( k^{}_{1}, k^{}_{2}) &=& 4
  D_{YBB} g^{}_{Y} g^{\,2}_{B} \, \frac{a^{}_{n}}{2} \,  \epsilon^{\lambda \nu \alpha \beta} 
k_{1 \alpha} k_{2 \beta} \\
k^{}_{2 \nu } {\bf V} ^{\lambda \mu \nu }_{YBB} \left(  k^{}_{1}, k^{}_{2}  \right) &=& 
4  D_{YBB} g^{}_{Y} g^{\,2}_{B} \, \frac{a^{}_{n}}{2} \,  \epsilon^{\lambda \mu \alpha \beta} 
k_{2 \alpha} k_{1 \beta}     \\
k^{}_{\lambda } {\bf V}^{ \lambda \mu \nu }_{YBB} ( k^{}_{1}, k^{}_{2}) &=& 0, 
\eeqn
where the chiral current Y has to be conserved so to render the 1 loop effective action gauge invariant. 
Introducing a symmetric distribution of the anomaly, in the BBB case the analogous equations are

\beqn
k^{}_{1\mu} {\bf V}^{\lambda \mu \nu}_{BBB}( k^{}_{1}, k^{}_{2}) &=& 4
  D_{BBB}  g^{\,3}_{B} \, \frac{a^{}_{n}}{3} \,   \epsilon^{\lambda \nu \alpha \beta} 
k_{1 \alpha} k_{2 \beta} \\
k^{}_{2 \nu } {\bf V} ^{\lambda \mu \nu }_{BBB} \left(  k^{}_{1}, k^{}_{2}  \right) &=& 
4  D_{BBB}  g^{\,3}_{B}  \, \frac{a^{}_{n}}{3} \,    \epsilon^{\lambda \mu \alpha \beta} 
k_{2 \alpha} k_{1 \beta}     \\
k^{}_{\lambda } {\bf V}^{ \lambda \mu \nu }_{BBB} ( k^{}_{1}, k^{}_{2}) &=& 4
 D^{}_{BBB} g^{\,3}_{B}  \, \frac{a^{}_{n}}{3} \,    \epsilon^{\mu \nu \alpha \beta} k_{1 \alpha} k_{2 \beta} , 
\eeqn

A study of the issue of the gauge dependence in these types of models can be found in \cite{CIM1}. Clearly, 
in our case, this study 
is more involved, but the cancelations of the 
gauge dependendent terms in specific classes of diagrams can be performed both in the 
exact phase and in the broken phase, similarly to the discussion presented in our companion work, 
having re-expressed the fields in the basis of the mass eigenstates. The approach that we follow is then clear: we worry about the cancelation of the anomalies in the exact phase, having performed a minimal gauge fixing to remove the B mixing with the axion $b$, then we rotate the fields and re-parameterize the lagrangean around the non trivial vacuum of the potential. We will see in the next sections that with this simple procedure we can easily discuss simple basic processes involving neutral and charged currents exploiting the invariance of the effective action under re-parameterizations of the fields.

%
%
\section{ The neutral currents sector in the MLSOM}
%
In this section we move toward the phenomenological analysis of a typical process which exhibits 
the new trilinear gauge interactions at 1-loop level. As we have mentioned in the introduction, our goal 
here is to characterize this analysis at a more formal level, leaving to future work a numerical study. It should be clear, however, from the discussion presented in this and in the next sections, how to proceed in a more general case. The theory is well-defined and consistent so that we 
can foresee accurate studies of its predictions for applications at the LHC in the future.  

We proceeed with our illustration starting from the definition of the neutral current in the model, which is given by 
\beqn
- {\mathcal L}_{NC} = \overline{\psi}^{}_{f} \gamma^\mu \left[ g^{}_{2} W^{3}_{\mu} T^{\,3} +  g^{}_{Y} Y A^{Y}_{\mu} 
+ g^{}_{B} Y^{}_{B} B^{}_{\mu}  \right]  \psi^{}_{f},
\eeqn
that we express in the two basis, the basis of the interaction eigenstates and of the mass eigenstates. 
Clearly in the  interaction basis the bosonic operator in the covariant derivative becomes
\beqn
\mathcal{F} &\equiv & g^{}_{2} W^{3}_{\mu} T^{3} +  g^{}_{Y} Y A^{Y}_{\mu} + g^{}_{B} Y^{}_{B} B^{}_{\mu} \nonumber \\
&=& g^{}_{Z} Q^{}_{Z} Z^{}_{\mu} 
+ g^{}_{Z^\prime} Q^{}_{Z^\prime} Z^{\prime}_{\mu} + e\, Q A^{\gamma}_{\mu}, 
 \eeqn
where $Q = T^{3} + Y$.
%
%
%
The rotation in the photon basis gives
\beqn
W^{3}_{\mu} &=& O^{A}_{W_{3} \gamma} A^{\gamma}_{\mu} + O^{A}_{W_{3} Z} Z_{\mu} + O^{A}_{W_{3} Z^\prime} Z^{\prime}_{\mu}  \\
A^{Y}_{\mu} &=& O^{A}_{Y \gamma} A^{\gamma}_{\mu} + O^{A}_{Y Z} Z_{\mu} + O^{A}_{Y Z^\prime} Z^{\prime}_{\mu}  \\
B_{\mu} &=& O^{A}_{B Z} Z_{\mu} + O^{A}_{B Z^\prime} Z^{\prime}_{\mu}  \\
\eeqn
and performing the rotation on $\mathcal{F}$  we obtain
\beqn
\mathcal{F}&=&  A^{\gamma}_{\mu} \left[ g^{}_{2} O^{A}_{W_{3} \gamma} T^{3} + g^{}_{Y} O^{A}_{Y \gamma} Y  \right] 
+  Z_{\mu} \left[   g^{}_{2} O^{A}_{W_{3} Z} T^{3} + g^{}_{Y} O^{A}_{Y Z} Y + g^{}_{B} O^{A}_{BZ} Y^{}_{B} \right]   \nonumber\\
&&+  Z^{\prime}_{\mu} \left[  g^{}_{2} O^{A}_{W_{3} Z^\prime} T^{3} + g^{}_{Y} O^{A}_{Y Z^\prime} Y 
+ g^{}_{B} O^{A}_{BZ^\prime} Y^{}_{B} \right],
\eeqn
where the electromagnetic current can be written in the usual way
\beqn
 A^{\gamma}_{\mu} \left[ g^{}_{2} O^{A}_{W_{3} \gamma} T^{3} + g^{}_{Y} O^{A}_{Y \gamma} Y  \right]  = 
e A^{\gamma}_{\mu} Q,
\eeqn
with the definition of the electric charge as
\beqn
e= g_{2} O^{A}_{W_{3} \gamma} = g_{Y} O^{A}_{Y \gamma} = \frac{ g^{}_{Y} g^{}_{2} }{ \sqrt{ g^{2}_{Y} + g^{2}_{2}}}. 
\eeqn
Similarly for the neutral Z current we obtain
\beqn
&& Z_{\mu} \left[  g^{}_{2} O^{A}_{W_{3} Z} T^{3} + g^{}_{Y} O^{A}_{Y Z} Y + g^{}_{B} O^{A}_{BZ} Y_{B} \right]  \nonumber\\
&=&  Z_{\mu} \left[ T^{3} ( g^{}_{2} O^{A}_{W_{3} Z} - g^{}_{Y} O^{A}_{YZ}) + g^{}_{Y} O^{A}_{YZ} Q 
+ g^{}_{B} O^{A}_{BZ} Y_{B} \right] \nonumber\\
&=&  Z_{\mu} g^{}_{Z}   \left[ T^{3} +   \frac{g^{}_{Y} O^{A}_{YZ}}{ g^{}_{2} O^{A}_{W_{3} Z} - g^{}_{Y} O^{A}_{YZ} } Q 
+ \frac{g^{}_{B} O^{A}_{BZ}}{ g^{}_{2} O^{A}_{W_{3} Z} - g^{}_{Y} O^{A}_{YZ} }  Y_{B}  \right],
\eeqn
where we have defined
\beqn
g_{Z} = g^{}_{2} O^{A}_{W_{3} Z} - g^{}_{Y} O^{A}_{YZ} \simeq  \, g = \frac{g^{}_{2}}{\cos \theta^{}_{W}}. 
\eeqn
We can easily work out the structure of the covariant derivative interaction 
applied on a left-handed or on a right-handed fermion.
For this reason it is convenient to introduce some notation. We define
\beqn
&&\mu^{Z}_{Q} = \frac{g^{}_{Y} O^{A}_{YZ}}{g_{Z}} \simeq - \sin^{2} \theta^{}_{W},  \\
&& \mu^{Z}_{B} = \frac{g^{}_{B} O^{A}_{BZ}}{g^{}_{Z}} \simeq  \frac{g^{}_{B}}{2} \epsilon^{}_{1} \mbox{\,\,\,\,\,so that\,\,} 
\lim_{M^{}_{1} \rightarrow \infty}  \mu^{Z}_{B}  = 0, 
\eeqn 
and similarly for the $Z^{\prime}$ neutral current 
\beqn
g_{Z^\prime} = g^{}_{2} O^{A}_{W_{3} Z^\prime} - g^{}_{Y} O^{A}_{YZ^\prime}, \qquad
\mu^{Z^{\prime}}_{Q} = \frac{g^{}_{Y} O^{A}_{YZ^\prime}}{g^{}_{Z^\prime}},  \qquad \mu^{Z^\prime}_{B} 
= \frac{g^{}_{B} O^{A}_{B Z^\prime}}{g^{}_{Z^\prime}}. 
\eeqn
We can easily identify the generators in the (Z, ${Z^\prime}$, $A_{\gamma}$) basis. These are given by
\beqn
\hat{Q}^{}_{Z} &=& \hat{Q}^{R}_{Z}   + \hat{Q}^{L}_{Z} = T^{3 L} + \mu^{Z}_{Q} Q^{L} 
+  \mu^{Z}_{B} Y^{L}_{B} +  \mu^{Z}_{Q} Q^{R} +  \mu^{Z}_{B} Y^{R}_{B}   \nonumber\\
\hat{Q}^{}_{Z^\prime} &=& \hat{Q}^{R}_{Z^\prime}  + \hat{Q}^{L}_{Z^\prime} = T^{3 L} + \mu^{Z^\prime}_{Q} Q^{L} 
+  \mu^{Z^\prime}_{B} Y^{L}_{B}  + \mu^{Z^\prime}_{Q} Q^{R} +  \mu^{Z^\prime}_{B} Y^{R}_{B}  \nonumber\\
\hat{Q}  &=& \hat{Q}_{L} + \hat{Q}_{R}
\eeqn
which will be denoted as $ Q^{}_{\, \overline p} = ( \hat{Q}, \hat{Q}_{Z}, \hat{Q}_{Z^\prime})$.
To express a given correlator, say $\langle Z A_{\gamma} A_{\gamma}  \rangle$ in the $(W_{3}, A_{Y}, B)$ basis we proceed as follows.
We denote with $ Q^{}_{\, \overline p} = ( \hat{Q}, \hat{Q}_{Z}, \hat{Q}_{Z^\prime})$ the generators in the photon basis 
$( A_{\gamma}, Z, Z^\prime)$ and with $g^{}_{\, \overline p}= (e, g^{}_{Z}, g^{}_{Z'})$ the corresponding couplings. Similarly, $ Q^{}_{p} = (T^{3}, Y, Y_{B})$ are 
the generators in the interaction basis 
$(W_{3}, A_{Y}, B)$ and $g^{}_{p} = (g^{}_{2}, g^{}_{Y}, g^{}_{B})$ the corresponding couplings, so that 
\beqn
- {\mathcal L}_{NC} &=& \overline{\psi} \gamma^{\mu} \left[  g^{}_{Z} \hat{Q}_{Z} Z_{\mu} 
+ g^{}_{Z^{\prime}} \hat{Q}_{Z^{\prime}} Z^{\prime}_{\mu} + e \, \hat{Q} A^{\gamma}_{\mu}  \right] \psi    \nonumber\\
&=&   \overline{\psi} \gamma^{\mu} \left[ g^{}_{2} T^{\,3} W^{\,3}_{\mu} 
+ g^{}_{Y} Y A^{Y}_{\mu}  + g^{}_{B} Y^{}_{B} B_{\mu}  \right] \psi.
\eeqn

\section{The $Z  \gamma \gamma$ vertex in the Standard Model}

Before coming to the computation of this vertex in the MLSOM we first start reviewing its structure in the SM. 

\begin{figure}[t]
{\centering \resizebox*{11cm}{!}{\rotatebox{0}
{\includegraphics{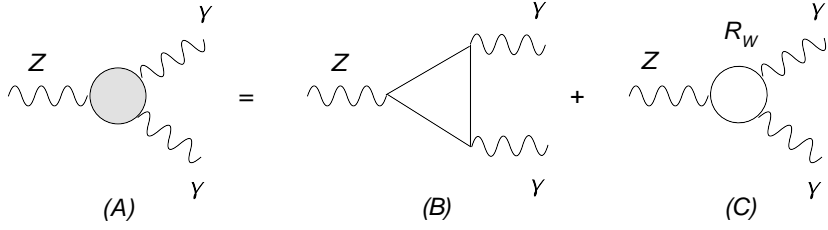}}}\par}
\caption{The $Z \gamma \gamma$ vertex to lowest order in the Standard Model, with the anomalous contributions and the remaining weak corrections shown separately. }
\label{ZGGcontributions}
\end{figure}

We show in Fig. \ref{ZGGcontributions} the $Z \gamma \gamma$ vertex in the SM, where we have separated
the QED contributions from the remaining corrections $R_{W}$. This vertex vanishes at all orders 
when all the three lines 
are on-shell, due to the Landau-Yang theorem. A direct prook of this property for the fermionic 1-loop corrections has been included in an appendix, where we show the on-shell vanishing of the vertex.

\begin{figure}[t]
{\centering \resizebox*{15cm}{!}{\rotatebox{0}
{\includegraphics{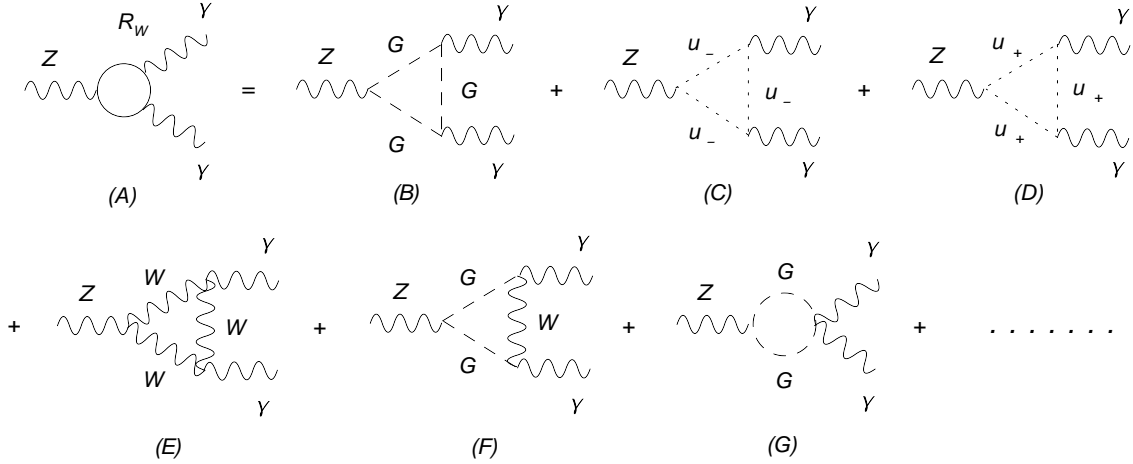}}}\par}
\caption{Some typical electroweak corrections, involving the charged Goldstones (here 
denoted by $G$, ghosts contributions ($u_\pm$) and W exchanges. }
\label{electroweak}
\end{figure}

\begin{figure}[t]
{\centering \resizebox*{6cm}{!}{\rotatebox{0}
{\includegraphics{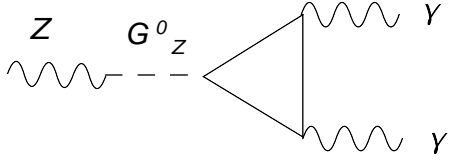}}}\par}
\caption{$Z-G^0_Z$ mixing in the broken phase in the SM. }
\label{gauge_mixing}
\end{figure}

\begin{figure}[t]
{\centering \resizebox*{10cm}{!}{\rotatebox{0}
{\includegraphics{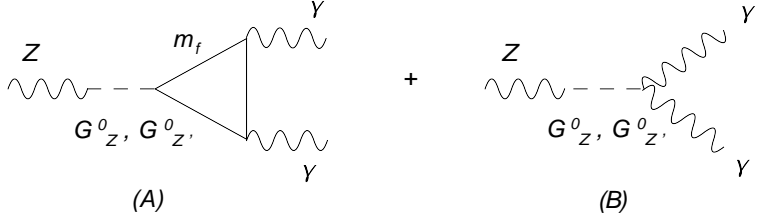}}}\par}
\caption{Same as in Fig.~\ref{gauge_mixing} but for the MLSOM}
\label{rotated_mixing_gauge}
\end{figure}

\begin{figure}[t]
{\centering \resizebox*{14cm}{!}{\rotatebox{0}
{\includegraphics{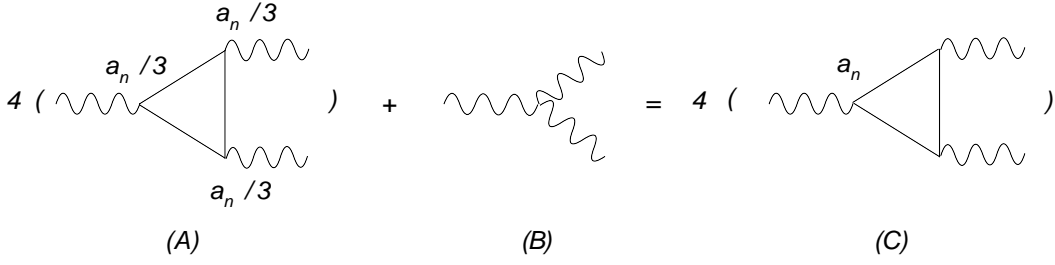}}}\par}
\caption{Re-distribution of the anomaly via the CS counterterm}
\label{anomaly_organization}
\end{figure}

The QED contribution contains the fermionic triangle diagrams (direct plus exchanged) and the contributions in $R_W$ include all the remaining ones at 1-loop level. 
In this case the separation between the pure QED contributions 
(due to the 2 fermionic diagrams) and the remaining  
corrections, which are separately gauge invariant on the photon lines, is rather straightforward, 
though this is not the case, in general, for more complicated electroweak amplitudes. 
Specifically, as shown in Fig. \ref{electroweak}, $R_{W}$, contains ghosts, 
goldstones and all other exchanges.
An exhaustive computation  of all these contributions 
is not needed for the scope of this discussion and will be left for future work. We have omitted diagrams of the type shown in Figs.~\ref{gauge_mixing},\ref{rotated_mixing_gauge}. These 
are removed by working in the $R_\xi$ gauge for the Z boson. 
Notice, however, that even without a gauge fixing these decouple 
from the anomaly diagrams in the massless fermion limit since the Goldstone does not couple to massless fermions. 
In Fig.~\ref{anomaly_organization} we show how the anomaly is re-distributed in an AAA diagram  
by a CS interaction, generating an AVV vertex. 

To appreciate the role played by the anomaly in this vertex we perform a direct 
computation of the two anomaly diagrams and include the fermionic mass terms. 
A direct computation gives



\beqn
G^{\rho \nu \mu} (k, k_1, k_2) &=& - \frac{e^2 g}{\cos \theta_{W} } \sum_{f} g^{f}_{A} Q^{2}_{f} \int \frac{d^{4} p}{(2 \pi)^{4}}
\, Tr\left( \frac{1}{{\slashed p} - m_f} \gamma^{\rho} \gamma^{5} \frac{1}{ {\slashed p} - {\slashed k} - m_f } \gamma^{\nu}
\frac{1}{ { \slashed p} - {\slashed k_1} - m_f } \gamma^{\mu}  \right)    \nonumber\\
&&+ \, (k_1 \rightarrow k_2, \mu \rightarrow \nu).
\eeqn
which can be cast in the form 
\beqn
G^{\rho \nu \mu} (k, k_1, k_2) &=& - \frac{e^2 g}{ 2  \pi^2 \cos{ \theta_W} } \sum_f g^{f}_{A} Q^{2}_{f} \int^{1}_{0} 
d x_1  \int^{1 - x_1 }_{0} d x_2 \,   \nonumber\\
&& \cdot \, \frac{1}{\Delta} \, \Big[ \epsilon^{\rho \nu \mu \alpha } 
(1 - x_1 - x_2)(x_2 k_1 - x_1 k_2)_{\beta} ( k^{\beta}_{2} k_{1 \alpha} + k^{\beta}_{1} k_{2 \alpha } )    \nonumber\\
&&  + (1 - x_1 - x_2 ) (  \epsilon^{\alpha \rho \beta \nu } k_{1 \alpha} k_{2 \beta} ( x_2 k^{\mu}_{1} -  x_1 k^{\mu}_{2} )  
+ (\mu \rightarrow \nu ) )   \nonumber\\
&&  + \epsilon^{\alpha \nu \beta \mu } k_{1 \alpha} k_{2 \beta } 
( x_2 ( x_2 -  x_1 - 1 )  k^{\rho}_{1} - x_1 ( x_2 -  x_1 + 1 ) k^{\rho}_{2} )  \Big],
\label{cast}
\eeqn
where
\beqn
\Delta = m^{2}_{f} + x_2 (x_2 -1) k^{2}_{1} + x_1 (x_1 - 1 ) k^{2}_{2} - 2 x_1 x_2 \, k_1 \cdot k_2,
\eeqn
and 
we have introducing the $g_{Z,A}^f$ and $g_{Z,V}^f$ couplings of the Z with 
\beq
g_{Z,A}^{f} = \frac{1}{2} T_3^f,  \qquad g_{Z,V}^{f} = \frac{1}{2} T_3^f - Q_f \sin^2\theta_W.
\eeq
This form of the amplitude is obtained 
if we use the standard Rosenberg definition of the anomalous diagrams and it agrees with \cite{Boud}. In this case 
the Ward identities on the photon lines are defining conditions for the vertex. 
Naturally, with the standard fermion multiplet assignment the anomaly vanishes since
\beqn
\sum_{f} g^{f}_{A} Q^{2}_{f} = 0. 
\eeqn 
Because of the anomaly cancelation, the fermionic vertex is zero 
also off-shell, 
if the masses of all the fermions in each generation are degenerate, in particular if they are massless.
Notice that this is not a consequence of the Landau-Yang theorem.
 
Let us now move to the Ward identity on the Z line. A direct computation gives
\beqn
k_{\rho} \, G^{\rho \nu \mu}  &=&  (k_1 + k_2 )_{\rho} \,  G^{\,  \rho \nu \mu }     \nonumber\\
&=& \frac{ e^2 g }{ \pi^{2} \cos \theta_{W}}  \sum_{f}  g^{f}_{A}  Q^{2}_{f} \,  \epsilon^{\, \nu \mu \alpha \beta } 
k_{1 \alpha} k_{2 \beta } \, \left[  \frac{1}{2} - m^{2}_{f} \int^{1}_{0} d x_1 \int^{1 - x_1 }_{0} 
d x_2  \,  \frac{1}{ \Delta }    \right]. 
\label{ABJ_anomaly}
\eeqn
The presence of a mass-dependent term on the right hand side of (\ref{ABJ_anomaly}) constitutes a break-down of axial 
current conservation for massive fermions, as expected.
\subsection{The $Z  \gamma \gamma$ vertex in anomalous abelian models: the Higgs-St\"uckelberg phase} 
The presence of anomalous generators in a given vertex renders some trilinear interactions non-vanishing also for massless fermions. In fact, 
as we have shown in the previous section, in the SM the anomalous triangle diagrams vanish if we neglect the masses of all the fermions, and this occurs both on-shell and off-shell. The only left over corrections 
are related to the fermion mass and these will also vanish (off-shell) if all the 
fermions of a given generation are mass degenerate.  
The on-shell vanishing of the same vertices is a consequence of the 
structure of the amplitude, as we show in the appendix. 
The extraction of the contribution of the anomalous generators in the trilinear vertices can be obtained starting from the 1-particle irreducible effective action, written in the basis of the interaction eigenstates, and performing the rotation of the trilinear interaction that 
project onto the $Z \gamma \gamma$ vertex.

In order to appreciate the differences between the SM result and the analogous one in the anomalous extensions that we are considering, we start by observing that only in the St\"uckelberg phase ($M_1\neq 0$ and $v_u=v_d=0$) the anomaly-free 
traces vanish,   
\ba
\langle Y Y Y \rangle \,g_Y^3 \, Tr [Q_Y^3]&=& 0 \nonumber \\
\langle Y W_3 W_3 \rangle \, g_Y g_2^2 \,Tr [Q_Y T^3 T^3]&=& 0\,,
\ea
because of charge assignment. A similar result is valid also in the 
HS phase if the Yukawa couplings are neglected. Coming to extract the 
$Z \gamma \gamma$ vertex we rotate the anomalous diagrams of the effective action into the mass eigenstates, being careful to separate the massless from the massive fermion contributions. 

Hence, we split the $\langle YYY\rangle$ vertex into its chiral contributions
and performing the rotation of the fields we get the following contributions
\ba
&&\frac{1}{3!}\, \langle Y Y Y \rangle \,g_Y^3 \, Tr [Q_Y^3]=\nonumber\\
&&\hspace{2.5cm}\sum_f\left[g_Y^3\frac{1}{8}(Q_{Y,f}^{L})^3\langle LLL \rangle^{\lambda\mu\nu}
+g_Y^3\frac{1}{8}(Q_{Y,f}^{R})^3\langle RRR \rangle^{\lambda\mu\nu}
\right.\nonumber\\
&&\hspace{2.5cm}\left.+g_Y^3\frac{1}{8}Q_{Y,f}^{L}(Q_{Y,f}^{R})^2 \langle LRR \rangle^{\lambda\mu\nu}
+g_Y^3\frac{1}{8}Q_{Y,f}^{L}Q_{Y,f}^{R}Q_{Y,f}^{L}\langle LRL \rangle^{\lambda\mu\nu}
\right.\nonumber\\
&&\hspace{2.5cm}\left.+g_Y^3\frac{1}{8}(Q_{Y,f}^{L})^2 Q_{Y,f}^{R}\langle LLR \rangle^{\lambda\mu\nu}
+g_Y^3\frac{1}{8}Q_{Y,f}^{R}(Q_{Y,f}^{L})^2 \langle RLL \rangle^{\lambda\mu\nu}
\right.\nonumber\\
&&\hspace{2.5cm}\left.+g_Y^3\frac{1}{8}Q_{Y,f}^{R}Q_{Y,f}^{L}Q_{Y,f}^{R}\langle RLR \rangle^{\lambda\mu\nu}
+g_Y^3\frac{1}{8}(Q_{Y,f}^{R})^2 Q_{Y,f}^{L}\langle RRL \rangle^{\lambda\mu\nu}
\right]
Z^{\lambda} A_{\g}^{\mu}A_{\g}^{\nu}\frac{1}{3!} R^{YYY}
+\dots \nonumber\\
\ea
where the dots indicate all the other projections of the type $ZZ\g,Z^{\prime}\g\g$ etc. 
Here $\langle LLL\rangle$, $\langle RLR\rangle$ etc., indicate 
the (clockwise) insertion of $L/R$ chiral projectors on the $\lambda \mu\nu$ vertices of the anomaly diagrams.  

For the $\langle YWW\rangle$ vertex the structure is more simple because the generator
associated to $W_3$ is left-chiral
\ba
&&\frac{1}{2!}\, \langle Y W W \rangle \,g_Y g_2^2 \, Tr [Q_Y (T^{3})^2]=
\sum_f\left[g_Y g_2^2\frac{1}{8}Q_{Y,f}^{L}(T^{3}_{L,f})^2\langle LLL \rangle^{\lambda\mu\nu}
\right.\nonumber\\
&&\left.\hspace{2.5cm}
+g_Y g_2^2\frac{1}{8}Q_{Y,f}^{R}(T^{3}_{L,f})^2\langle RLL \rangle^{\lambda\mu\nu}
\right]Z^{\lambda} A_{\g}^{\mu}A_{\g}^{\nu}\frac{1}{2!} R^{YWW}
+\dots \nonumber\\
\ea
The $\langle BYY\rangle$ vertex works in same way of $\langle YYY\rangle$
\ba
&&\frac{1}{2!}\, \langle B Y Y \rangle \,g_B g_Y^2 \, Tr [Q_B Q_Y^2]=\nonumber\\
&&\hspace{1.5cm}\sum_f\left[g_B g_Y^2\frac{1}{8}Q_{B,f}^{L}(Q_{Y,f}^{L})^2\langle LLL \rangle^{\lambda\mu\nu}
+g_B g_Y^2\frac{1}{8}Q_{B,f}^{R}(Q_{Y,f}^{R})^2\langle RRR \rangle^{\lambda\mu\nu}
\right.\nonumber\\
&&\hspace{1.5cm}\left.+g_B g_Y^2\frac{1}{8}Q_{B,f}^{L}(Q_{Y,f}^{R})^2 \langle LRR \rangle^{\lambda\mu\nu}
+g_B g_Y^2\frac{1}{8}Q_{B,f}^{L}Q_{Y,f}^{R}Q_{Y,f}^{L}\langle LRL \rangle^{\lambda\mu\nu}
\right.\nonumber\\
&&\hspace{1.5cm}\left.+g_B g_Y^2\frac{1}{8}Q_{B,f}^{L}Q_{Y,f}^{L}Q_{Y,f}^{R}\langle LLR \rangle^{\lambda\mu\nu}
+g_B g_Y^2\frac{1}{8}Q_{Y,f}^{R}(Q_{Y,f}^{L})^2 \langle RLL \rangle^{\lambda\mu\nu}
\right.\nonumber\\
&&\hspace{1.5cm}\left.+g_B g_Y^2\frac{1}{8}Q_{B,f}^{R}Q_{Y,f}^{L}Q_{Y,f}^{R}\langle RLR \rangle^{\lambda\mu\nu}
+g_B g_Y^2\frac{1}{8}Q_{B,f}^{R} Q_{Y,f}^{R}Q_{Y,f}^{L}\langle RRL \rangle^{\lambda\mu\nu}
\right]Z^{\lambda} A_{\g}^{\mu}A_{\g}^{\nu}\frac{1}{2!} R^{BYY}
+\dots \nonumber\\
\ea
Finally, the $\langle BWW\rangle$ vertex is similar to $\langle YWW\rangle$
\ba
&&\frac{1}{2!}\, \langle B W W \rangle \,g_Y g_2^2 \, Tr [Q_B (T^{3})^2]=
\sum_f\left[g_B g_2^2\frac{1}{8}Q_{B,f}^{L}(T^{3}_{L,f})^2\langle LLL \rangle^{\lambda\mu\nu}
\right.\nonumber\\
&&\left.\hspace{5cm}
+g_B g_2^2\frac{1}{8}Q_{B,f}^{R}(T^{3}_{L,f})^2\langle RLL \rangle^{\lambda\mu\nu}
\right]Z^{\lambda} A_{\g}^{\mu}A_{\g}^{\nu}\frac{1}{2!} R^{BWW}
+\dots \nonumber\\
\ea
where we have defined
\ba
&&R^{YYY}=3\left[(O^{A\,T})_{22}(O^{A\,T})_{21}^{2}\right]\nonumber\\
&&R^{YWW}=\left[2(O^{A\,T})_{11}(O^{A\,T})_{12}(O^{A\,T})_{21}+(O^{A\,T})_{11}^{2}(O^{A\,T})_{22}\right]\nonumber\\
&&R^{BYY}=(O^{A\,T})_{21}^{2}(O^{A\,T})_{32}\nonumber\\
&&R^{BWW}=\left[(O^{A\,T})_{11}^{2}(O^{A\,T})_{32}\right]\,.
\ea
which are the product of rotation matrices that project the anomalous effective 
action from the interaction eigenstate basis over the $Z,\gamma$ gauge bosons.

We have expressed the generators in their chiral basis, and their mixing is due to mass insertions over each
fermion line in the loop. The ellypsis refers to additional contributions which do not project
on the vertex that we are interested in but which are present
in the analysis of the remaining neutral vertices, $ZZ\g,Z^{\prime}\g\g$ etc. The notation
$O^{AT} $ indicates the transposed of the rotation matrix from the interaction to the mass eigenstates.
To obtain the final expression of the amplitude in the interaction eigenstate basis one can easily observe that in the helicity conserving amplitudes $\langle LLL \rangle$ and $\langle RRR\rangle$  
the mass dependence in the fermion loops  is all contained in the denominators of the propagators, not in the Dirac traces. The only diagrams that contain a mass dependence at the numerators are those involving chirality flips 
($\langle LLR\rangle,  \langle RRL \rangle$) which contribute with terms 
proportional to $m_f^2$. These terms contribute only to the invariant amplitudes 
$A_1$ and $A_2$ of the Rosenberg representation \cite{CIM1} and, although finite, they disappear once we impose a Ward identity on the two photon lines, as requested by CVC for the two photons.   A similar result is 
valid for the SM, as one can easily figure out from  Eq. (\ref{cast}). Therefore, the amplitudes can be expressed just in terms of $LLL$ and $RRR$ correlators, and since the mass dependence is at the denominators of the propagators,  one can easily show the relation 
\beq
\langle LLL\rangle =-\langle RRR\rangle
\eeq
valid for any fermion mass $m_f$. Defining
$\langle LLL\rangle\equiv\Delta^{\lambda\mu\nu}_{LLL}(m_f\neq 0)$, we can express the only independent chiral graph as sum of two contributions
\ba
\Delta^{\lambda\mu\nu}_{LLL}(m_f\neq 0)=\Delta^{\lambda\mu\nu}_{LLL}(0)+\Delta^{\lambda\mu\nu}_{LLL}(m_f)
\ea
where we define
\ba
&&\Delta^{\lambda\mu\nu}_{LLL}(0)\equiv\Delta^{\lambda\mu\nu}_{LLL}(m_f=0)\nonumber\\
&&\Delta^{\lambda\mu\nu}_{LLL}(m_f)\equiv\Delta^{\lambda\mu\nu}_{LLL}(m_f\neq 0)-\Delta^{\lambda\mu\nu}_{LLL}(m_f=0).
\ea
Also, one can verify quite easily that
\ba
\Delta^{\lambda\mu\nu}_{LLL}( 0) &=&\Delta^{\lambda\mu\nu}_{AVV}(0)+\Delta^{\lambda\mu\nu}_{VAV}(0)+
\Delta^{\lambda\mu\nu}_{VVA}(0)+\Delta^{\lambda\mu\nu}_{AAA}(0)\nonumber \\
&&=4 \Delta^{\lambda\mu\nu}_{AAA}(0).\nonumber\\
\ea

A second contribution to the effective action comes from the 1-loop
counterterms containing generalized CS terms. There are two ways to express these counterterms:
either as separate 3-linear interactions or as
modifications of the two invariant amplitudes of the Rosenberg
parameterization $A_1, A_2$. These amplitude depend linearly on the momenta of the vertex \cite{CIM1}.
For instance we use

\ba
\Delta_{AAA}(0)-\frac{a_n}{3}\varepsilon^{\lambda\mu\nu\alpha}(k_{1\alpha}-k_{2\alpha})=\Delta_{AVV}(0),
\ea
which allows to absorb completely the CS term, giving conserved 
$Y/W_3$ currents in the interaction eigenstate basis. In this case we move 
from a symmetric distribution of the anomaly  in the $AAA$ diagram, to an 
$AVV$ diagram. 
These currents interpolate with the vector-like vertices (V) of the AVV graph. 

Notice that once the anomaly is moved from any vertex involving a $Y/W_3$ current to a vertex with a $B$ current, it is then canceled by the GS interaction.
The extension of this analysis to the complete $m_f$-dependent case 
for $\Delta_{LLL}(m_f\neq 0)$ is  
quite straightforward.   
In fact,  after some re-arrangements of the $Z\g\g$ amplitude, we are left with the following contributions in the physical basis in the broken phase
\ba
&&\langle Z\g\g\rangle|_{m_f\neq 0}=
\frac{1}{4}\sum_f\Delta_{AVV}^{\lambda\mu\nu}(m_f\neq 0)
\left[g_Y^3\theta_f^{YYY}R^{YYY}+ g_Y g_2^2\theta_f^{YWW}R^{YWW}
\right.\nonumber\\
&&\hspace{2cm}\left.
+g_B g_Y^2\theta_f^{BYY}R^{BYY}+ g_B g_2^2\theta_f^{BWW}R^{BWW}\right]Z^{\lambda}A_{\g}^{\mu}A_{\g}^{\nu}
\label{final}
\ea
where we have defined the anomalous chiral asymmetries as
\ba
&&\theta^{BYY}_{f}=\left[Q_{B,f}^{L} (Q_{Y,f}^{L})^2-Q_{B,f}^{R} (Q_{Y,f}^{R})^2\right]
\nonumber\\
&&\theta^{BWW}_{f}=Q_{B,f}^{L}(T^{3}_{L,f})^2.
\ea
The conditions of gauge invariance force the coefficients in front of the CS terms to be
\ba
&&D_{BYY}=\frac{1}{8}\sum_f\theta^{BYY}_{f}
\nonumber\\
&&D_{BWW}=\frac{1}{8}\sum_f\theta^{BWW}_{f},\, 
\ea
which have been absorbed and do not appear explicitly, while the SM chiral asymmetries are defined as
\ba
&&\theta^{YYY}_{f}=\left[(Q_{Y,f}^{L})^3-(Q_{Y,f}^{R})^3\right]
\nonumber\\
&&\theta^{YWW}_{f}=Q_{Y,f}^{L}(T^{3}_{L,f})^2,
\ea
and the triangle $\Delta_{AVV}(m_f\neq 0)$ is given as in (\ref{cast}).
Notice that Eq. (\ref{final}) is in complete agreement with the SM result 
shown in (\ref{cast}), obtained by removing the contributions proportional to the $B$ gauge bosons  and setting the chiral asymmetries of $Y$ and $W_3$ to zero. 
In particular, if the gauge bosons are not anomalous and in the 
chiral limit ($m_f=0$ or  $m_f=m$) 
this trilinear amplitude vanishes. 
 
As we have already pointed out, the amplitude for the $\langle Z\gamma\gamma\rangle$ process 
is espressed in terms of 6 invariant amplitudes that can be easily computed and take the form

\beqa
\Delta^{\lambda \mu\nu}_{AVV} &=& {A}_1(k_1,k_2) \epsilon[k_1,\mu,\nu,\la] +
{A}_2(k_1,k_2)\epsilon[k_2,\mu,\nu,\la]
+{A}_3(k_1,k_2) \epsilon[k_1,k_2,\mu,\la]{k_1}^{\nu}\nonumber \\
&& + {A}_4(k_1,k_2) \epsilon[k_1,k_2,\mu,\la]{k_2}^{\nu} +
{A}_5(k_1,k_2) \epsilon[k_1,k_2,\nu,\la]k_1^\mu +
{A}_6(k_1,k_2)\epsilon[k_1,k_2,\nu,\la]k_2^\mu, \nonumber \\
\label{Ros}
\eeqa
with
\beqa
A_1(k_1,k_2)&=&k_1\cdot k_2 A_3(k_1,k_2) + k_2^2 A_4(k_1,k_2) \nonumber \\ 
A_2(k_1,k_2) &=& -A_1(k_2,k_1) \nonumber \\
A_5(k_1,k_2) &=& -A_4(k_2,k_1) \nonumber \\
A_6(k_1,k_2) &=& -A_3(k_2,k_1). \nonumber \\
\eeqa
Also $A_1(k_1,k_2)=A_1(k_2,k_1)$ as one can easily check by a direct computation. 
We obtain 
\beqa
&&A_3(k_1,k_2)= -\frac{1}{2}\int_0^1 dx \int_0^{1-x} dy \frac{x y}{y (1-y) k_1^2 + 
x(1-x) k_2^2 + 2 x y \,k_1\cdot k_2 - m_f^2}  \nonumber \\
&& A_4(k_1,k_2)= \frac{1}{2}\int_0^1 dx \int_0^{1-x} dy \frac{x (1-x)}{y (1-y) k_1^2 + 
x(1-x) k_2^2 + 2 x y \,k_1\cdot k_2 - m_f^2}  \nonumber \\
\eeqa
The computation of these integrals can be done analytically and the various regions 
$0<  s< 4 m_f^2 $, $m_f>> \sqrt{s}/2$, and $m_f \to 0$  can be studied in detail.
In the case of both photons on-shell, for instance, and $s > 4 m_f^2 $ we obtain 
\beqa
&&A_3(k_1,k_2) = \frac{1}{2 s} - \frac{m_f^2}{s} \makebox{Li}_2\left(\frac{2}{1 - 
\sqrt{1- 4 m_f^2/s}}\right) - \frac{m_f^2}{s} \makebox{Li}_2\left(\frac{2}{1 + 
\sqrt{1- 4 m_f^2/s}}\right) \nonumber \\
&& A_4(k_1,k_2)= 
-\frac{1}{s} + \frac{\sqrt{1- 4 m_f^2/s}}{s}\makebox{ArcTanh}\left(\frac{1}
{\sqrt{1- 4 m_f^2/s}}\right)\eeqa
Notice that the case in which the two photons are on-shell and light fermions are running in the 
loop, then the evaluation of the integral requires particular care because 
of infrared effects which render the parameteric integrals ill-defined. 
The situation is similar to the case of the coupling of the axial anomaly to on-shell 
gluons in spin physics \cite{CollinsMueller}, when the correct isolation of the massless quarks contributions 
is carried out by moving off-shell on the external lines and then performing the $m_f\to 0$ limit.
%
\subsection{ $q \bar{q} \to \gamma \gamma$ with an intermediate Z}
In this section we are going to describe the role played by the new anomaly cancelation 
mechanism in simple processes which can eventually be studied with accuracy at a hadron collider such as the LHC. A numerical analysis of processes involving neutral 
currents can be performed along the lines of \cite{CCG} and we hope to return to this point in the near future. Here we intend to discuss briefly some of the phenomenological 
implications which might be of interest. 
Since the anomaly is canceled by a combination of Chern-Simons 
and Green-Schwarz contributions, the study of a specific process, such as 
$Z \to \gamma \gamma $, which differs from the SM prediction, requires, in general, 
a combined analysis both of the gauge sector and of the scalar sector.

\begin{figure}[t]
{\centering \resizebox*{14cm}{!}{\rotatebox{0}
{\includegraphics{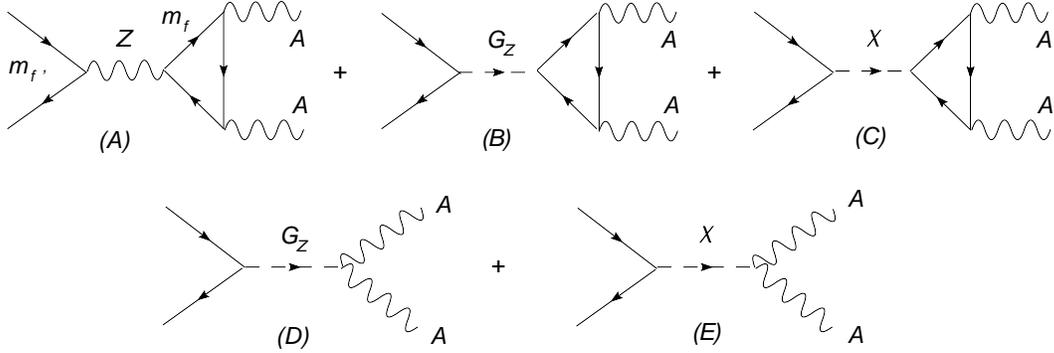}}}\par}
\caption{Two photon processes initiated by a $q \bar{q}$ annihilation with a $Z$ exchange.}
\label{ampiezza3}
\end{figure}

We start from the case of a quark-antiquark annihilation mediated by a Z that later undergoes a decay into two photons. At leading order this process is at parton level described by the annihilations of a valence quark $q$ 
and a sea antiquark $\bar{q}$ from the two incoming hadrons, both of them collinear and massless.
In Fig.~(\ref{ampiezza3}) we have depicted all the diagrams by which the process can take place to lowest 
order. Radiative corrections from the initial state are accurately known up to next-to-next-to-leading order, and are universal, being the same of the Drell-Yan cross section.
In this respect, precise QCD predictions for the rates are available, for instance around the Z resonance \cite{CCG}.  

In the SM, gauge invariance of the process requires both a $Z$ gauge boson exchange 
and the exchange of the corresponding goldstone $G_Z$, which involves 
diagrams (A) and (B). In the MLSOM a direct Green-Schwarz coupling to the photon (which is gauge dependent) 
is accompanied by a gauge independent axion exchange. If the incoming 
quark-antiquark pair is massless, then the Goldstone has no coupling to the incoming fermion pair, and 
therefore (B) is absent, while gauge invariance is trivially satisfied because of the massless condition 
on the fermion pair of the initial state. In this case only diagram (A) 
is relevant. Diagram (B) may also be set to vanish, for instance in suitable gauges, such as the unitary gauge. Notice also that the triangle diagrams have a dependence on $m_f$, the mass of the fermion in the loop, 
and show two contributions: a first contribution which is proportional to the anomaly (mass independent) and a correction term which depends on $m_f$. 

As we have shown above, the first contribution, which involves an off-shell vertex, is absent in the SM,
while it is non vanishing in the MLSOM. In both cases, on the other hand, we have $m_f$ dependent contributions. 
It is then clear that in the SM the largest contribution to the process comes from the 
top quark circulating in the triangle diagram, the amplitude being essentially proportional only to the heavy top mass. On the $Z$ resonance and for on-shell photons, the cross section vanishes in both cases, as we have explained, in agreement with the Landau-Yang theorem. We have checked these properties explicitly, but they 
hold independently of the perturbative order at which they are analyzed, being based on the Bose 
symmetry of the two photons. The cross section, therefore, has a dip at $Q=M_Z$, where it vanishes, and where $Q^2$ is the virtuality of the intermediate s-channel exchange.

\begin{figure}[t]
{\centering \resizebox*{15cm}{!}{\rotatebox{0}
{\includegraphics{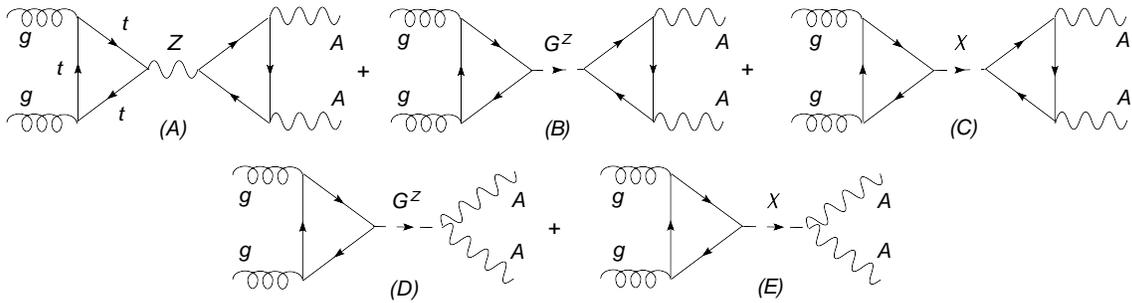}}}\par}
\caption{Gluon fusion contribution to double-photon production. Shown are also the scalar 
exchanges (B) and (D) that restore gauge invariance and the axi-Higgs exchange (E).}
\label{gluoncorretto}
\end{figure}

An alternative scenario is 
to search for neutral exchanges initiated by gluon-gluon fusion. In this case we replace the annihilation pair with a triangle loop (the process is similar to Higgs production via gluon fusion), as shown in Fig. 
\ref{gluoncorretto}. 
As in the decay mechanism discussed above, the production 
mechanism in the SM and in the MLSOM are again different. In fact, 
in the MLSOM there is a massless contribution appearing already at the massless fermion level, which is absent in the SM. The production mechanism by gluon fusion has some special features as well.
In ggZ production and $Z\gamma\gamma$ decay, the relevant diagrams are (A) and (B) since we need the 
exchange of a $G_Z$ to obtain gauge invariance. 
As we probe smaller values of the Bjorken variable $x$,  the gluon density raises, and the process 
becomes sizable. 
On the other hand, in a pp collider, although the quark annihilation channel 
is suppressed since the antiquark density 
is smaller than in a $p \bar{p}$ collision, 
this channel still remains rather significant. We have also shown in this figure one of the scalar channels, due to the exchange of a axi-Higgs. 

\begin{figure}[t]
{\centering \resizebox*{12cm}{!}{\rotatebox{0}
{\includegraphics{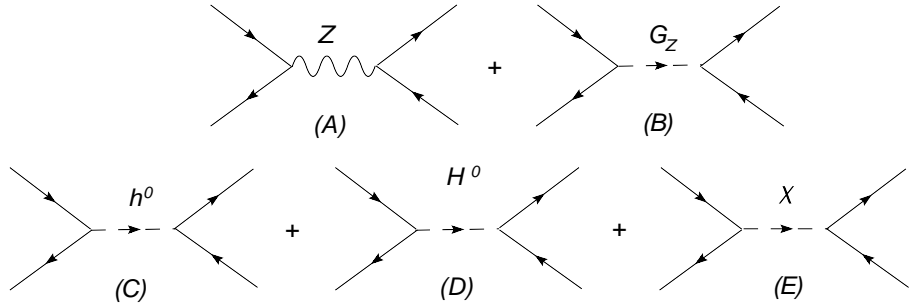}}}\par}
\caption{The $q \bar{q}$ annihilation channel (A,B). 
Scalar exchanges in the neutral sector 
involving the two Higsses and the Axi-Higgs (C,D,E). }
\label{ampiezza2}
\end{figure}

Other channels such as those shown in Fig.~\ref{ampiezza2} can also be studied, these involve a 
lepton pair in the final state, and their radiative corrections also show the appearance of a triangle 
vertex. This is the classical Drell-Yan process, that we will briefly describe below. 
In this case, both the total cross section and the rapidity distributions of the lepton pair and/or an analysis 
of the charge asymmetry in s-channel exchanges of W's would be of major 
interest in order to disentangle the anomaly inflow. At the moment, errors on the parton 
distributions and scale dependences induce indeterminations which, just for the QCD 
background, are around $4 \% $ \cite{CCG}, as shown in a high precision study. It is expected, 
however, that the statistical accuracy on the $Z$ resonance at the LHC is going to be a factor 100 better. 
In fact this is a case in which the experiment can do better than the theory. 

\subsection{Isolation of the massless limit: the $Z^*\to \gamma^* \gamma^* $ amplitude}
%
%
The isolation of the massless from the massive contributions can be analized in the case of 
resolved photons in the final state. As we have already mentioned in the prompt photon case the amplitude, on the Z resonance, vanishes because of Bose symmetry and angular momentum conservation. We can, 
however, be on the $Z$ resonance and produce one or two off-shell photons that undergo fragmentation. 
Needless to say, these contributions are small. However, the separation of the massless 
from the massive case is well defined. One can increase the rates by asking just for 1 single resolved photon and 1 prompt photon. Rates for this process in pp-collisions have been determined in \cite{CorianoGordon}. We start from the case of off-shell external 
photons of virtuality $s_1$ and $s_2$  and an off-shell Z $(Z^*)$.     
Following \cite{Kuhn}, we introduce the total vertex $V^{\lambda\mu\nu}
(k_1,k_2,m_f)$, which contains both the massive $m_f$ dependence (corresponding to the triangle amplitude $\Delta^{\lambda\mu\nu }$. Its massless counterpart ${\bf V}^{\lambda\mu\nu}(0)\equiv V(k_1,k_2,m_f=0)$, obtained by sending the fermion mass to zero. 
The Rosenberg vertex and the V vertex are trivially related by a Schoutens 
transformation, moving the $\lambda$ index from the Levi-Civita tensor to the momenta 
of the photons
\beqn
&&V_{\lambda \mu \nu} (k_1, k_2, m_{f})    \nonumber\\
&&  = A(k_1, k_2,m_f) \varepsilon[\lambda, \mu, \nu, k_2] s_1 
- A(k_2, k_1,m_f) \varepsilon[\lambda, \mu , \nu, k_1] s_2 + A(k_1, k_2,m_f) \varepsilon[\lambda, \nu, k_1, k_2] k_1^{\mu}  \nonumber\\
&&+ A(k_2, k_1,m_f) \varepsilon[\lambda, \mu, k_2, k_1] k^{\nu}_{2} - B(k_1, k_2,m_f) \varepsilon[\mu, \nu, k_1, k_2] k^{\lambda}
\eeqn
with $k - k_1 - k_2 = 0$ and $s_i = k_i^2 \,\, (i=1,2)$, and  
\beqn
A (k_1, k_2, m_f) &=& \frac{1}{\lambda} \left[ - \frac{1}{2}(s -s_1 + s_2) - \left( \frac{1}{2}(s + s_2) 
+ ( 6/{\lambda}) s s_1 s_2  \right) \Delta_{\#1} \right.   \nonumber\\
&& + s_2  \left[ \frac{1}{2} 
- ( 3/{\lambda}) s (s - s_1 - s_2) \right] \Delta_{\#2}   \nonumber\\
&&   \left. + \left[ s s_2 + (m_f^2 + (3/{\lambda})s s_1 s_2)(s 
- s_1 + s_2)   \right] C_{\#0} \right]
\label{Aeq}
\eeqn
\beqn
B(k_1, k_2, m_f) &=& \frac{1}{\lambda} \left[ \frac{1}{2} (s - s_1 - s_2) + s_1 \left[ \frac{1}{2} 
+ (3/{\lambda}) s_2 (s + s_1 - s_2)  \right]   \Delta_{\#1}   \right.   \nonumber\\
&&+ s_2 \left[ \frac{1}{2} 
+ (3/{\lambda}) s_1 (s - s_1 + s_2)  \right] \Delta_{\#2}     \nonumber\\
&&   \left. + \left[ s_1 s_2 
- (m_f^2 + (3/{\lambda}) s s_1 s_2)(s - s_1 - s_2)  \right] C_{\#0} \right]
\label{Beq}
\eeqn
with
\beq
\lambda = \lambda(s, s_1, s_2),  \nonumber\\
\eeq
being the usual Mandelstam function and
where the analytic expressions for $\Delta_{\#i}$ and $C_{\# 0}$ are given by
\beqn
\Delta_{\#i} &=& a_i \, \mbox{ln} \frac{a_i + 1}{a_i -1} - a_3 \, \mbox{ln} \frac{a_3 +1}{a_3 - 1} , \,\,\,\mbox{(i=1,2)}   \nonumber\\
C_{\#0} &=& \frac{1}{\sqrt{\lambda} } \sum^{3}_{i=1} \left[ \mbox{ Li}_{2} \left( \frac{b_i -1}{a_i + b_i} \right)
- \mbox{Li}_{2} \left( \frac{- b_i -1}{a_i - b_i} \right) + \mbox{Li}_{2} \left( \frac{- b_i +1}{a_i - b_i} \right) 
- \mbox{Li}_{2} \left( \frac{b_i + 1}{a_i + b_i} \right)  \right],     \nonumber\\
\eeqn
and
\beqn
&&t_i = - s_i - i \epsilon,     \qquad    a_i = \sqrt{1 + (2 m_f)^{2} /t_i}, \,\,\mbox{(i=1,2,3)},   \nonumber\\
&&\lambda = \lambda(t_1, t_2, t_3),  \qquad   b_1 = (t_1 - t_2 -t_3)/\sqrt{\lambda} \,\,\,\,\mbox{or cyclic}
\eeqn
For $m_f=0$ the two expressions above become 
\beqn
\Delta_{\#i} &=& \mbox{ln}(t_i/t_3), \,\,\,(i=1,2),  \nonumber\\
C_{\# 0} &=& (1/\sqrt{\lambda})  \Biggr[ 2  \Biggr( \zeta (2) - \mbox{Li}_{2}(x_1) - \mbox{Li}_{2}(x_2) 
+ \mbox{Li}_{2}  \left(  \frac{1}{x_3}  \right) \Biggr) + \mbox{ln} x_1 \mbox{ln}x_2 \Biggr]
\eeqn
with
\beqn 
x_i = \frac{(b_i + 1)}{(b_i - 1)}, \,\,\, (i = 1,2,3).
\eeqn
These can be inserted into (\ref{Aeq}) and (\ref{Beq}) together with $m_f=0$ to 
generate the corresponding ${\bf V}^{\lambda\mu\nu}(0)$ vertex needed for the computation of the massless contributions to the amplitude. 

With these notations we clearly have 
\beqa
\Delta^{\lambda\mu\nu} &=& V^{\lambda\mu\nu}(k_1,k_2,m_f) \nonumber \\
\Delta^{\lambda\mu\nu}(0) &=& {\bf V}^{\lambda\mu\nu}(k_1,k_2) \nonumber \\
\Delta^{\lambda\mu\nu}(m_f) &=& V^{\lambda\mu\nu}(k_1,k_2,m_f)
 - {\bf V}^{\lambda\mu\nu}(k_1,k_2). \nonumber \\
\eeqa
%
\subsection{Extension to $Z\to \gamma^* \gamma$}
%
To isolate the contribution to the decay on the resonance, we keep one of the two photons off-shell 
(resolved). We choose $s_1=0$, and $s_2$ virtual. We denote by $\Gamma^{\lambda\mu\nu}$ the 
corresponding vertex in this special kinematical configuration. The Z boson is on-shell.  
In this case at 1-loop the result simplifies considerably \cite{Hagiwara}
\beqn
\Gamma_{\lambda \mu \nu} &=& F_{2} (s_2 \epsilon[\lambda, \mu, \nu, k_1] + k^{\nu}_{2} \epsilon[\lambda, \mu, k_1, k_2] ),
\label{ampiezza_Zgg}
\eeqn
with $F_2$ expressed as a Feynman parametric integral 
\beqn
F_{2} =  \frac{1}{{2 \pi^{2}}}\int^{1}_{0} dz_{1} dz_{2} dz_{3} \delta(1 - z_{1} - z_{2} - z_{3}) 
\frac{- z_{2} z_{3}}{ m_f^{2} - z_{2}z_{3} s_2 - z_{1} z_{3} M_Z^2}.
\eeqn
Setting $F_2\equiv - F(z,r_f)$
where $f(z,r)$ is a dimensionless function of
\beqn
z= s_2/M^{2}_{Z}, \qquad r_f=m^{2}_{Z}/4m_{f}^2,
\eeqn
and for vanishing $m_f$ ($r_{f} = M^{2}_{Z}/4m^{2}_{f} \rightarrow \infty$), 
the corresponding massless contribution is expressed as $F(z,\infty)$
with, in general
\beqn
F(z,r_f) = \frac{1}{4(1 - z)^{2}} (I(r_f z,r_f) - I(r_f,r_f) + 1 - z ),
\eeqn
where
\beqn
I(x,r_f) &=& 2 \sqrt{\frac{x - 1}{x^{-}}} \mbox{ln}(\sqrt{-x} + \sqrt{1-x})
 - \frac{1}{r_f} (\mbox{ln}(\sqrt{-x} + \sqrt{1 -x}))^{2}  \,\,\,\,\,\,\,\,  \mbox{for \,x $<$ 0}  \nonumber\\
 &=& 2 \sqrt{\frac{1-x}{x}} \sin^{-1} \sqrt{x} + \frac{1}{r_f} (\sin^{-1} \sqrt{x})^{2} \,\,\,\,\,\,\,\mbox{for 0$<$\,x $<$ 1}  \nonumber\\
&=& 2 \sqrt{\frac{x-1}{x}} \left( \ln(\sqrt{x} + \sqrt{x-1}) - \frac{i \pi}{2}  \right) 
- \frac{1}{r_f} \left( \ln(\sqrt{x} + \sqrt{x-1} ) - \frac{i \pi}{2}  \right)^{2},    \nonumber\\
&&  \,\,  \mbox{for \,x $>$ 1}.
\eeqn
The $m_f=0$ contribution is obtained in the $r_f \rightarrow + \infty$ limit,
\beqn
F(z,\infty) &=& \frac{1}{4(1 - z)^{2}} (\ln z + 1 - z)  \,\,\,\,\,\,\,\,\,\,\,\,\,\,\,\,\,\,  \mbox{for \, z $>$ 0},  \nonumber\\
&=& \frac{1}{4(1 - z)^{2}} (\ln |z| + i \pi + 1 - z)  \,\,\,\,\,\,\,\mbox{for \, z $<$ 0}.
\eeqn
In these notations, the infinite fermion mass limit 
($m_{f} \rightarrow \infty$ or $r \rightarrow 0$), 
gives $F(z,0)= 0$ and we find 
\beqa
\Delta^{\lambda\mu\nu} &=& \Gamma^{\lambda\mu\nu}=F(z,r_f) \nonumber \\
\Delta^{\lambda\mu\nu}(0) &=& { \Gamma}^{\lambda\mu\nu}(0)=F(z,\infty) \nonumber \\
\Delta^{\lambda\mu\nu}(m_f) &=& \Gamma^{\lambda\mu\nu}- \Gamma^{\lambda\mu\nu}(m_f)=F(z,r_f) 
-F(z,\infty),
\eeqa
which can be used for a numerical evaluation. 
The decay rate for the process is given by 
\beq
\Gamma(Z\to \gamma^*\gamma)=
\frac{1}{4 M_Z}\int d^4 k_1 d^4 k_2 \delta(k_1^2)\,\delta(k_2^2 - Q_*^2) 
|{\cal M}_{Z\to \gamma\gamma^*}|^2 \,2 (\pi)^4 \delta( k - k_1 - k_2),
\eeq
where 
\beqa
|{\cal M}_{Z\to \gamma\gamma^*}|^2 &=& - A^{\lambda \mu\nu}_{Z\to \gamma\gamma^*}
\Pi_Z^{\lambda \lambda'}A^{\lambda' \mu\nu'}_{Z\to \gamma\gamma^*}\Pi_{Q^*}^{\nu \nu'}
\nonumber \\
\Pi_Z^{\lambda \lambda'} &=& - g^{\lambda \lambda'} + \frac{k^{\lambda}k^{\lambda'}}{M_Z^2} \nonumber \\
\Pi_{Q^*}^{\nu \nu' } &=& - g^{\lambda \lambda'} + \frac{k^{\lambda}k^{\lambda'}}{Q_*^2}. 
\eeqa
We have indicated with $Q_*$ the virtuality of the photon. A complete evaluation 
of this expression, to be of practical interest, would need the fragmentation functions of the photon 
(see \cite{CorianoGordon} for an example). A detailed analysis of these rates will be presented elsewhere. However, we will briefly summarize the main points 
involved in the analysis of this and similar processes at the LHC, where the decay rate 
is folded with the (NLO/NNLO) contribution from the initial state using QCD factorization.

Probably one of the best way to search for neutral current interactions in hadronic collisions at the LHC is in 
lepton pair production via the Drell-Yan mechanism. 
 QCD corrections are known for this process up to O($\alpha_s^2$) (next-to-next-to-leading order, NNLO), 
which can be folded with the 
NNLO evolution of the parton distributions to provide accurate determinations of the hadronic pp cross 
sections at the 4 $\%$ level of accuracy \cite{CCG}. The same computation for Drell-Yan can be used 
to analize the $pp\to Z\to \gamma \gamma^*$ process 
since the $W_V$ (hadronic) part of the process is universal, with $W_V $ defined below. An appropriate (and very useful) way to analyze this process would be to perform this study defining the invariant mass distribution

\beq
\frac{d \sigma}{d Q^2}= \tau \sigma_{Z\to \gamma^*\gamma}
(Q^2,M_V^2)\,W_V(\tau, Q^2) 
\eeq
where $\tau=Q^2/S$, 
which is separated into a pointlike contribution $\sigma_{Z\to \gamma \gamma^*}$
\beq
\sigma_V(Q^2,M_V^2)=\frac{\pi \alpha}{4 M_Z \sin\theta_W^2 \cos\theta_W^2 N_c}
\frac{\Gamma(Z\to \gamma\gamma^*)}{(Q^2 - M_Z^2)^2 + M_Z^2 \Gamma_Z^2}.
\eeq
and a hadronic structure functions $W_Z$.
This is defined via the integral over parton distributions and coefficient functions 
$\Delta_{ij}$ 
\beq
W_Z(Q^2,M_Z^2)=\int_0^1 d x_1\, \int_0^1 d x_2 \int_0^1 d x \delta( \tau - x x_1 x_2) 
P D^V_{ij}(x_1,x_2,\mu_f^2) \Delta_{ij}(x, Q^2,\mu_f^2)
\eeq
where $\mu_f$ is the factorization scale. The choice $\mu_f=Q$, with Q the invariant mass of the 
$\gamma\gamma^*$ pair , removes the $\log(Q/M)$ for the computation of the coefficient functions, which is, anyhow, 
arbitrary.
The non-singlet coefficient functions are given by 
\beqa
\Delta^{(0)}_{q \bar{q}} &=& \delta(1-x) \nonumber \\
\Delta^{(1)}_{q \bar{q}} &=& \frac{\alpha_S(M_V^2)}{4 \pi}C_F\left[ \delta(1-x)( 8 \zeta(2) - 16) + 
16 \left(\frac{\log(1-x)}{1-x}\right)_+ \right.\nonumber \\
&&\left. \qquad \qquad \quad \qquad -8 (1 + x)\log(1 -x) - 4\frac{1 + x^2}{1 -x} \log x\right]
\eeqa
with $C_F=(N_c^2-1)/(2 N_c)$ and the ``+'' distribution is defined by
\beq
 \left(\frac{\log(1-x)}{1-x}\right)_+= \theta( 1-x) \frac{\log(1-x)}{1-x} -
\delta(1-x)\int_0^{1-\delta} dx \, \frac{\log(1-x)}{1-x},
\eeq
while at NLO appears also a q-g sector 
\beq
\Delta^{(1)}_{q g} = \frac{\alpha_S(M_V^2)}{4 \pi}T_F\left[ 2(1 + 2 x^2 -2 x)\log 
\left( \frac{(1 -x)^2}{x}\right) + 1 - 7 x^2 + 6 x\right].
\eeq
Other sectors do not appear at this order.
Explicitly one gets 
\beqa
W_Z(Q^2,M_Z^2) &=&\sum_i \int_0^1 d x_1\, \int_0^1 d x_2 
\int_0^1 d x \delta( \tau - x x_1 x_2) \nonumber \\
 && \times \Big\{ \bigg(q_i(x_1,\mu_f^2)\bar{q_i}(x_2,\mu_f^2) +  \bar{q_i}(x_1,\mu_f^2)
{q_i}(x_2,\mu_f^2)\bigg) \Delta_{q\bar{q}}(x, Q^2,\mu_f^2) \nonumber \\
&&\qquad  + \bigg(q_i(x_1,\mu_f^2){g}(x_2,\mu_f^2) +  
{q_i}(x_2,\mu_f^2){g}(x_1,\mu_f^2)\bigg) \Delta_{q{g}}(x, Q^2,\mu_f^2)\Big\} \nonumber \\
\eeqa
where the sum is over the quark flavours. The identification of the generalized mechanism of anomaly cancelation requires 
that this description be extended to NNLO, which is now a realistic 
possibility. It involves a slight modification of the NNLO hard scatterings known at 
this time and an explicit computation is in progress.

%
%
%
%
%
%

\section{Conclusions} 
%
We have presented a study of a model inspired by the structure encountered
in a typical string theory derivation of the Standard Model. In particular we have focused our investigation 
on the characterization of the effective action and worked out its expression 
in the context of an extension containing one additional anomalous $U(1)$. Our analysis specializes and, at the same time, extends a previous study of models belonging to this class.
The results that we have presented are generic for models where the St\"{u}ckelberg and the Higgs mechanism are combined and where an effective abelian anomalous interaction is present. 
Our analysis has then turned toward the study of simple processes mediated by neutral current exchanges, and we have focused, specifically, on one of them, the one involving the $Z \gamma \gamma$ vertex. In particular our findings clearly show that 
new massless contributions are presented at 1-loop level when anomalous generators are involved in the fermionic triangle diagrams and the interplay between massless and massive fermion effects is modified respect to the SM case.
The typical processes considered in our analysis deserve a special attention, 
given the forthcoming experiments at the LHC, since they may provide a way to determine whether anomaly effects 
are present in some specific reactions. Other similar processes, involving the entire neutral sector should be considered, though the two-photon signal 
is probably the most interesting one phenomenologically.

Given the high statistical precision 
($.05 \%$ and below on the Z peak, for 
10 $fb^{-1}$ of integrated luminosity) which can be easily obtained at the LHC, there are realistic chances to prove or disprove theories of these types. Concerning the possibility of discovering extra anomalous $Z'$, although there are stringent 
upper bounds on their mixing(s) with the Z gauge boson, it is of outmost importance to bring 
this type of analysis even closer to the experimental test by studying 
in more detail the peculiarities of anomalous gauge interactions for both the neutral and the charged sectors along the lines developed in this work. 
This analysis is in progress and we hope to report on it in the near future.

\centerline{\bf Acknowledgements} 
{\em We dedicate this work to the memory of Hidenaga Yamagishi, remembering his remarkable scientific 
talent, his outstanding human qualities and his unique and inspirational love for physics.} 

We thank Elias Kiritsis for having brought the topics discussed in this work to our attention and Theodore Tomaras, Marco Roncadelli, Marco Guzzi, 
Roberta Armillis, Andrea Spirito and Antonio Quintavalle for discussions. 
The work of C.C. was supported in part by the European Union through the Marie Curie Research and Training Network ``Universenet'' (MRTN-CT-2006-035863). 
He thanks the Theory Group at the University of Liverpool and in particular Alon Faraggi for discussions and for the kind hospitality. N.I. was partially supported by the European contract MRTN-CT-2004-512194.

\section{Appendix. A Summary on the single anomalous $U(1)$ model.}
We summarize in this appendix some results concerning the model with a 
single anomalous $U(1)$ discussed in the main sections. These results 
specialize and simplify the general discussion of \cite{CIK} to which we refer 
for further details.
We will use the hypercharge values
%
\hskip 2cm
\begin{center}
\begin{tabular}{|c|c|c|c|c|c|c|}
\hline
$ f $ & $Q_{L}$ &  $ u_{R} $ &  $ d_{R} $ & $ L $ & $e_R$ & $\n_R$ \\
\hline \hline
$q^{}_Y$  &  $1/6$  & $2/3$  &  $-1/3$ & $-1/2$ & $-1$ & $0$\\ \hline
\end{tabular}
\end{center}
%
and general $U(1)^{}_{B}$ charge assignments   \\
%
\hskip 2cm
\begin{center}
\begin{tabular}{|c|c|c|c|c|c|c|}
\hline
$ f $ & $Q_{L}$ &  $ u_{R} $ &  $ d_{R} $ & $ L $ & $e_R$ & $\n_R$ \\
\hline \hline
$q^{}_B$  &  $q^{(Q_L)}_{B}$  & $q^{(u_R)}_{B}$  &  $q^{(d_R)}_{B}$ & $q^{(L)}_{B}$ & $q^{(e_R)}_{B}$ & $q^{(\nu_R)}_{B}$\\ \hline
\end{tabular}
\end{center}
The covariant derivatives act on the fermions $f_L,f_R$ as
\begin{eqnarray}
&& {\cal D}_{\mu}f^{}_{L} = \left(\partial_{\mu} +
i {\bf A}_{\mu} + i q_{l}^{(f_L)} g_{l}A_{l,\mu} \right)f^{}_{L}\nonumber \\
&& {\cal D}_{\mu}f^{}_{R} = \left(\partial_{\mu} +
i {\bf A}_{\mu} + i q_{l}^{(f_R)} g_{l}A_{l,\mu} \right)f^{}_{R}
\end{eqnarray}
with $l=Y,B$ abelian index, 
where ${\bf A}_{\mu}$ is a non-abelian Lie algebra element and write the lepton doublet as
\bea
L_{i} = \pmatrix { \nu^{}_{Li} \cr e^{}_{Li}}.
\eea
We will also use standard notations for the $SU(2)_W$ and $SU(3)_C$ gauge bosons 

\beqa
W_{\mu} &=& \frac{\sigma_i}{2} W^{i}_{\mu} = \tau_i W^{i}_{\mu}, \;\;\; \; \;  \mbox{with} \; \; \;\;   i = 1,2,3   \\
G_{\mu} &=& \frac{\lambda_a}{2} G^{\,a}_{\mu}  = T_a G^{\,a}_{\mu} \;\;\; \; \;  \mbox{with} \; \; \;\;   a = 1, 2,...,8
\eeqa
with the normalizations 

\beqn
Tr[ \tau^i \tau^j] = \frac{1}{2} \delta_{ij},    \qquad   Tr[ T^a T^b] = \frac{1}{2} \delta_{ab}.
\label{normalization}
\eeqn
The interaction lagrangean for the leptons becomes 
\beqa
 {\cal L}_{int}^{lep} &= &   \pmatrix{ \overline{{\nu}}_{Li} & \overline{e}_{Li}} \gamma^{\mu}
\left[ -  g^{}_2 \frac{\tau^a}{2} W_{\mu}^a  - g^{}_Y \, q^{(L)}_Y  A^Y_{\mu} 
- g^{}_B  q^{(L)}_B  B_{\mu} \right]  \pmatrix{\n_{Li} \cr e_{Li}}  + \nonumber\\
&& +  \; \overline{e}_{Ri} \; \gamma^{\mu} \left[
- g^{}_Y  q^{(e_{R})}_Y    A^Y_{\mu} -  g^{}_B  q^{(e_{R})}_B B_{\mu} \right]e_{Ri}\nonumber\\
&& +  \; \overline{\nu}_{Ri} \; {\gamma}^{\mu}\left[
- g^{}_Y  q^{({\n}_{R})}_Y  A^Y_{\mu} - g^{}_B  q^{(\n_{R})}_B   B_{\mu}\right]{\n}_{Ri}.
\eeqa
As usual we define the left-handed and right-handed currents 
\beqn
J^{L}_{\mu} = \frac{1}{2}(J_{\mu} - J^{5}_{\mu}),  \qquad  J^{R}_{\mu} = \frac{1}{2}(J_{\mu} + J^{5}_{\mu}), 
\qquad J_{\mu} = J^{R}_{\mu} + J^{L}_{\mu}, \qquad  J^{5}_{\mu} = J^{R}_{\mu} - J^{L}_{\mu}.
\eeqn
Writing the quark doublet as
\bea
Q^{}_{Li} = \pmatrix {u^{}_{Li}\cr d^{}_{Li}},
\eea
we obtain the interaction lagrangean
\beqa
 {\cal L}_{int}^{quarks} &= &   \pmatrix{ \overline{ u }^{}_{Li} & \overline{ d }^{}_{Li}} \gamma^{\mu}
\left[ - g^{}_3 \frac{\lambda^a}{2} G^{a}_{\mu}  -  g^{}_2 \frac{ \tau^i }{ 2 } W_{ \mu }^i  - g^{}_Y \, q^{(Q_L)}_Y  A^Y_{\mu} 
- g^{}_B  q^{(Q_L)}_B  B_{\mu} \right]  \pmatrix{ u^{}_{Li} \cr d^{}_{Li}}  + \nonumber\\
&& +  \; \overline{u}_{Ri} \; \gamma^{\mu} \left[
- g^{}_Y  q^{(u_{R})}_Y    A^Y_{\mu} - g^{}_B  q^{(u_{R})}_B B_{\mu} \right]  u^{}_{Ri}  \nonumber\\
&& +  \; \overline{d}^{}_{Ri} \; {\gamma}^{\mu}\left[
- g^{}_Y  q^{({d}_{R})}_Y  A^Y_{\mu} - g^{}_B  q^{(d_{R})}_B   B_{\mu}\right] d^{}_{Ri}.
\eeqa

As we have already mentioned in the introduction, we work with a 2-Higgs doublet model, and therefore 
we parameterize the Higgs fields in terms of 8 real degrees of freedom as
\beqa
H_u=\left(\begin{array}{c}
H_u^+\\
H_u^0
\end{array}\right) \qquad H_d = \left(\begin{array}{c}
H_d^+\\
H_d^0 \end{array}\right)
\eeqa
where $H_u^+$, $H_d^+$ and $H_u^0$, $H_d^0$ are complex fields. Specifically
\beq
H_u^+ =  \frac{H_{uR}^+ + i H_{uI}^+}{\sqrt{2}} ,\qquad
H_d^- =  \frac{H_{dR}^- + iH_{dI}^-}{\sqrt{2}} , \qquad
H_u^- = H_u^{+ *}, \qquad
H_d^+ = H_d^{- *}.
\eeq
Expanding around the vacuum we get for the uncharged components
\beq
H_u^0 =  v_u + \frac{H_{uR}^0 + i H_{uI}^0}{\sqrt{2}} , \qquad
H_d^0 =  v_d + \frac{H_{dR}^0 + iH_{dI}^0}{\sqrt{2}}. \label{Higgsneut}
\eeq
The Weinberg angle is defined via
$\cos\theta_W= g_2/g, \sin\theta_W= g_Y/g$, with
\bea
 g^2= g_Y^2 + g_2^2.
\eea
 We also define $\cos \beta=v_d/v$, $ \sin \beta=v_u/v$ and 
\bea
v^2=v_d^2 + v_u^2.
\eea  
The mass matrix in the mixing of the neutral gauge bosons is given by
\bea
{\mathcal L}_{mass} =  \left( W_3 \,\,\,\, A^Y \,\,\,\, B  \right){\bf M}^2    \left(\begin{array}{c}
W_3\\
A^Y\\
B \\
\end{array}   \right),
\eea
where   
\bea
{\bf M}^2 = {1\over 4} \pmatrix{
{g^{}_2}^{2} v^2 & - {g^{}_2} \, {g^{}_Y} v^2 &  - {g^{}_2} \,  x^{}_B \cr
 - {g^{}_2} \,{g^{}_Y} v^2 &  {g^{}_Y}^{2} v^2 & {g^{}_Y}  x^{}_B \cr
 -{g^{}_2} \, x^{}_B &{g^{}_Y}  x^{}_B  & 2 M_1^2 + N^{}_{BB}}
\label{massmatrix}
\eea
with
\bea
N^{}_{BB}=  \left( q_u^{B\,2} \,{v^{\,2}_u} + q_d^{B\,2} \,{v^{\,2}_d} \right)\, g_B^{\,2},
\eea
\bea
x^{}_B=  \left(q_u^B {v^{\,2}_u} + q_d^B {v^{\,2}_d}  \right)\, g^{}_B.
\eea
The orthonormalized mass squared eigenstates corresponding to this matrix are given by
\beqa
\left(
\begin{array}{c}
 O_{11}^{A} \\
 O_{12}^A \\
 O_{13}^A
\end{array}
\right)
=\left(
\begin{array}{c}
  \frac{ g_Y }{ \sqrt{ g_2^2 + g_Y^2 }} \\
  \frac{ g_2}{\sqrt{  g_2^2 + g_Y^2 }} \\
  0 
\end{array}
\right),
\eeqa
\beqa
\left(
\begin{array}{c}
 O_{21}^{A} \\
 O_{22}^A \\
 O_{23}^A
\end{array}
\right)
=   \left(
\begin{array}{c}
   \frac{ g_2 \left( 2 M_1^2 - g^2 v^2 + N_{BB} 
+ \sqrt{  \left(  2 M_1^2 - g^2
   v^2 + N_{BB} \right)^2 + 4 g^2
   x_{B}^2}   \right)} 
{   g^2 x_B   \sqrt{  4  +  \frac{g^2}{g^4 x_B^2} 
\left( 2 M_{1}^2 - g^2 v^2 + N_{BB} 
+ \sqrt{ \left( 2 M_{1}^2 - g^2 v^2 + N_{BB} \right)^2 + 4 g^2 x_{B}^2 } \right)^2 } }   \\
    - \frac{ g_Y  \left( 2 M_1^2 - g^2 v^2 + N_{BB} 
+ \sqrt{  \left(  2 M_1^2 - g^2
   v^2 + N_{BB} \right)^2 + 4 g^2
   x_{B}^2}   \right)} 
{   g^2 x_B   \sqrt{  4  +  \frac{g^2}{g^4 x_B^2} 
\left( 2 M_{1}^2 - g^2 v^2 + N_{BB} 
+ \sqrt{ \left( 2 M_{1}^2 - g^2 v^2 + N_{BB} \right)^2 + 4 g^2 x_{B}^2 } \right)^2 } }    \\
  \frac{2}{  \sqrt{  4  +  \frac{g^2}{g^4 x_B^2} 
\left( 2 M_{1}^2 - g^2 v^2 + N_{BB} 
+ \sqrt{ \left( 2 M_{1}^2 - g^2 v^2 + N_{BB} \right)^2 + 4 g^2 x_{B}^2 } \right)^2 } }  
\end{array}
\right).
\eeqa
One can see that these results reproduce the analogous relations of the SM  in the limit of very large $M_1$
\bea
\lim_{M_{1}\to\infty} O_{21}^{A} = \frac{g_2}{g}, \qquad  \lim_{M_{1}\to\infty} O_{22}^{A} = - \frac{g_Y}{g},  \nonumber\\
 O_{23}^{A}  \simeq \frac{g}{2} \frac{x_B}{M_{1}^2} \equiv  \frac{g}{2} \epsilon_1  \,\,\,
 \mbox{ so that} \,\,\, \lim_{M_{1}\to\infty} O_{23}^{A} = 0.   \nonumber
\eea
Similarly, for the other matrix elements of the rotation matrix $O^A $ we 
obtain
\beqa
\left(
\begin{array}{c}
 O_{31}^{A} \\
 O_{32}^A \\
 O_{33}^A
\end{array}
\right)
= \left(
\begin{array}{c}
 -  \frac{ g_2 \left( - 2 M_1^2 + g^2 v^2 - N_{BB} 
+ \sqrt{  \left(  2 M_1^2 - g^2
   v^2 + N_{BB} \right)^2 + 4 g^2
   x_{B}^2}   \right)} 
{   g^2 x_B   \sqrt{  4  +  \frac{g^2}{g^4 x_B^2} 
\left(- 2 M_{1}^2 + g^2 v^2 - N_{BB} 
+ \sqrt{ \left( 2 M_{1}^2 - g^2 v^2 + N_{BB} \right)^2 + 4 g^2 x_{B}^2 } \right)^2 } }   \\
     \frac{ g_Y  \left( - 2 M_1^2 + g^2 v^2 - N_{BB} 
+ \sqrt{  \left(  2 M_1^2 - g^2
   v^2 + N_{BB} \right)^2 + 4 g^2
   x_{B}^2}   \right)} 
{   g^2 x_B   \sqrt{  4  +  \frac{g^2}{g^4 x_B^2} 
\left( -2 M_{1}^2 + g^2 v^2 - N_{BB} 
+ \sqrt{ \left( 2 M_{1}^2 - g^2 v^2 + N_{BB} \right)^2 + 4 g^2 x_{B}^2 } \right)^2 } }    \\
  \frac{2}{  \sqrt{  4  +  \frac{g^2}{ g^4 x_B^2 } 
\left(- 2 M_{1}^2 + g^2 v^2 - N_{BB} 
+ \sqrt{ \left( 2 M_{1}^2 - g^2 v^2 + N_{BB} \right)^2 + 4 g^2 x_{B}^2 } \right)^2 } }  
\end{array}
\right),
\eeqa
whose asymptotic behavior is described by the limits 
\bea
O_{31}^{A}  \simeq -\frac{g^{}_2}{2} \frac{x^{}_B}{M^2_1} \equiv -\frac{g^{}_2}{2} {\epsilon_1} , 
\qquad  O_{32}^{A} \simeq \frac{g^{}_Y}{2} \frac{x^{}_B}{M^2_1} \equiv  \frac{g^{}_Y}{2}  \epsilon_1,
 \qquad   O_{33}^{A} \simeq  1,  
\eea
\bea
\lim_{M_{1}\to\infty} O_{31}^{A} = 0, \qquad  \lim_{M_{1}\to\infty} O_{32}^{A} = 0, \qquad   \lim_{M_{1}\to\infty} O_{33}^{A} = 1.
\eea
These mass-squared eigenstates correspond to one zero mass eigenvalue 
for the photon $A^{}_{\gamma}$, and two non-zero mass 
eigenvalues for the Z and for the ${Z}^{\prime}$ vector bosons, corresponding to the mass values
\bea
m_{Z}^2 &=&  \frac{1}{4} \left( 2 M_1^2 + g^2 v^2 + N^{}_{BB} 
- \sqrt{\left(2  M_{1}^2 - g^2 v^2 + N^{}_{BB} \right)^2 + 4
   g^2 x_{B}^2} \right)    \\
&\simeq&     \frac{g^2 v^2}{2} - \frac{1}{M_{1}^2} \frac{g^2 x_{B}^{2}}{4}
 + \frac{1}{M_{1}^4}\frac{g^2 x_{B}^2}{8 } (N^{}_{BB} - g^2 v^2) , \nonumber\\
\nonumber\\
 m_{{Z}^\prime}^2 &=&   \frac{1}{4} \left( 2 M_1^2 + g^2 v^2 + N^{}_{BB} 
+ \sqrt{\left(2  M_{1}^2 - g^2 v^2 + N^{}_{BB} \right)^2 + 4   g^2 x_{B}^2} \right)   \\
&\simeq&    M^{2}_{1} +  \frac{N^{}_{BB}}{2} . \nonumber
\eea
The mass of the $Z$ gauge boson gets corrected by terms 
of the order $v^{2}/M_1$, converging to the SM value as $M_1\to \infty$, 
with $M_1$ the St\"{u}ckelberg mass of the B gauge boson, 
the mass of the $Z'$ gauge boson can grow large with $M_1$. 

The physical gauge fields can be obtained from the rotation matrix $O^A$ 
\ba
\pmatrix{A_\g \cr Z \cr {{Z^\prime}}} =
O^A\, \pmatrix{W_3 \cr A^Y \cr B}  \label{OA}
\ea
which can be approximated at the first order as

\bea
O^A  \simeq  \pmatrix{
\frac{g^{}_Y}{g}           &     \frac{g^{}_2}{g}         &      0   \cr
\frac{g^{}_2}{g} + O(\epsilon_1^2)          &     -\frac{g^{}_Y}{g} + O(\epsilon_1^2) &      \frac{g}{2} \epsilon_1    \cr
-\frac{g^{}_2}{2}\epsilon_1     &     \frac{g^{}_Y}{2}\epsilon_1  &   1 + O(\epsilon_1^2) }   .  
\label{matrixO}
\eea
The mass squared matrix (\ref{massmatrix}) can be diagonalized as
\bea
 \Big( A_\g \,\,\,\, Z \,\,\,\, Z^\prime   \Big) \, O^A   {\bf M}^2  (O^A)^T  \left(\begin{array}{c}
A_\g\\
Z\\
Z^\prime \\
\end{array}   \right)  =   \Big( A_{\g} \,\,\,\,  Z \,\,\,\,  Z^\prime  \Big)   \pmatrix{
0           &     0         &  0   \cr
0           &    m_{Z}^2    & 0    \cr
0           &     0         &   m_{{Z}\prime}^2 }      \left(\begin{array}{c}
A_{\g}  \\
Z   \\
Z^\prime   \\
\end{array}   \right).
\eea
It is straightforward to verify that the rotation matrix $O^A$ satisfies the proper orthogonality relation
\bea
O^A (O^A)^T = 1. 
\eea
\subsection{Rotation matrix $O^\chi$ on the axi-Higgs}

This matrix is needed in order to rotate into the mass eigenstates of the CP odd sector, relating the axion $\chi$ and the two neutral Goldstones of this sector to the St\"uckelberg field $b$ and 
the CP odd phases of the two Higgs doublets

\beq
\pmatrix{{\rm Im}H_u^0\cr {\rm Im}H_d^0\cr  b \cr }=\; O^{\chi} \;
\pmatrix{\chi \cr G_1^{\,0} \cr G_2^{\,0} \cr }.
\label{rotunit}
\eeq
We refer to \cite{CIK} for a morre detailed discussion of 
the scalar sector of the model, 
where, in the presence of explicit phases (PQ breaking terms), 
the mass of the axion becomes massive from the massless case. 
The PQ symmetric contribution is given by

\bea
V_{PQ} = \sum_{a=u,d} \Bigl(  \mu_a^2  H_a^{\dagger} H_a + \l_{aa} (H_a^{\dagger} H_a)^2\Bigr)
-2\l_{ud}(H_u^{\dagger} H_u)(H_d^{\dagger} H_d)+2{\l^\prime_{ud}} |H_u^T\tau_2H_d|^2,
\eea
while the PQ breaking terms are 
\begin{eqnarray}
V_{\slash{P} \slash{Q}} &=&  b_{1} \, \left( H_u^{\dagger} H_d \, e^{-i  (q_u^B-q_d^B) \frac{b}{M_1}}  \right)
+ \lambda^{}_1 \left( H_u^{\dagger}H_d \,e^{-i  (q_u^B-q_d^B) \frac{b}{M_1}} \right)^2  \nonumber\\
&&+ \, \lambda^{}_2 \left( H_u^{\dagger}H_u \right) \left( H_u^{\dagger}H_d \,e^{-i  (q_u^B-q_d^B) \frac{b}{M_1}} \right)
+ \lambda^{}_3 \left( H_d^{\dagger}H_d \right) \left( H_u^{\dagger}H_d \,e^{-i (q_u^B-q_d^B)  \frac{b}{M_1}} \right) + c.c.
\nonumber\\ \label{PQbreak}
\end{eqnarray}
where $b^{}_{1}$ has mass squared dimension, while $\lambda^{}_{1}$, $\lambda^{}_{2}$, $\lambda^{}_{3}$ are dimensionless.

\bea
c^{}_{\chi} = 4 \left( 4 \lambda_1 +  \lambda_3 \cot \beta + \frac{ b_{1} }{ v^2 } \frac{ 2 }{ \sin 2\beta } 
+  \lambda_2  \tan\beta  \right),
\eea
and using $v_{d}=v \cos\beta,  \qquad   v_{u}=v \sin\beta$ together with
\bea
 \cot\beta=\frac{\cos\beta}{\sin\beta} = \frac{v_{d}}{v_{u}}, \qquad
\tan\beta =\frac{\sin\beta}{\cos\beta} = \frac{v_{u}}{v_{d}},
\eea

from the scalar potential \cite{CIK} one can extract the mass 
eigenvalues of the model for the sscalar sector. The mass matrix has 2 zero eigenvalues and one non-zero eigenvalue that corresponds to a physical
axion field, $ \chi$, with mass
\bea
m_{\chi}^2 =
-\frac{1}{2} \, c^{}_{ \chi} \, v^2  \left[ 1 + \left(  \frac{q_u^B-q_d^B}{M_1}\,
 \frac{v \, \sin{2\b} }{2} \right)^2 \right] = 
-\frac{1}{2} \, c^{}_{ \chi} \, v^2  \left[ 1 +  \frac{ ( q_u^B - q_d^B )^{2}}{M^{2}_{1}} \,
\frac{v^{2}_{u} v^{2}_{d} }{v^2}  \right].
\label{axionmass}
\eea
The mass of this state is positive if $c_{\chi} < 0$. Notice that the mass of the axi-Higgs is the result of two effects: 
the presence of the Higgs vevs and the presence of a PQ-breaking 
potential whose parameters can be small enough to drive the mass of 
this particle to be very light. We refer to \cite{CIM1} for a simple 
illustration of this effect in an abelian model. In the case of a single anomalous $U(1)$ $O^\chi$ can be simplified as shown below.

Introducing $N$ given by
\bea
N = \frac{1}{ \sqrt{ 1+ \frac{  ( q_u^B - q_d^B )^2 }{ M^{\,2}_1 }  \frac{ v_d^2 v_u^2 }{ v^2 } } } 
  = \frac{1}{ \sqrt{ 1+  \frac{ ( q_u^B - q_d^B )^2 }{ M^{\,2}_1 }  \frac{ v^2 \sin^2{2\beta} }{4}  }}   \label{normcoeff}
\eea
and defining
\bea
Q_1 = - \frac{ ( q_u^B - q_d^B )}{M_1} v_u  = - \frac{  (q_u^B-q_d^B)}{M_1} v \sin{\b},
\eea

\bea
N_1 = \frac{1}{\sqrt{ 1 + Q_1^2 }},
\eea

$O^{\chi}$ he following matrix
\beqa
O^{\chi} = \pmatrix{
-N\cos{\b} & \sin{\b} & {\overline N}_1 {\overline Q}_1 \cos{\b} \cr
 N\sin{\b} & \cos{\b} & -{\overline N}_1 {\overline Q}_1 \sin{\b} \cr
NQ_1 \cos{\b}& 0 & {\overline N}_1 },\nonumber\\
\eeqa
where we defined
\beq
{\overline Q}_1 = Q_1 \cos{\b}
\eeq
and
\beqa
{\overline N}_1 &=& \frac{1}{\sqrt{1+{\overline Q}_1^2}} = \frac{1}{\sqrt{ 1 +  Q_1^2 \cos^2\beta}}   \nonumber\\
&=&  \frac{1}{\sqrt{ 1 +  \frac{ (q_u^B  - q_d^B)^2 }{ M_1^{\,2} } \, v^2 \sin^2\beta \cos^2\beta}} 
=   \frac{1}{ \sqrt{ 1 + \frac{  (q_u^B  - q_d^B)^2 }{ M_1^{\,2} } \frac{ v_u^2 v_d^2}{v^2} }}.
\eeqa
One can see from (\ref{normcoeff}) that  ${\overline N}_1 = N$, and the 
explicit 
elements of the 3-by-3 rotation matrix $O^{\chi}$ can be written as
\bea
\left( O^{\chi}  \right)_{11} &=&  - \frac{1}{  \frac{ -  ( q_u^B - q_d^B ) }{M_1} v_u
   \sqrt{ \frac{M_1^{\,2} }{  ( q_u^B - q_d^B )^2 } \frac{ v^2 }{ v_u^2 v_d^2 } + 1 } }   \nonumber\\
   &=&     - \frac{1}{ v_u \, \frac{ v  }{v_u v_d}   } \, N      = -N\cos{\b}    \\
\left( O^{\chi}   \right)_{21}&=&    \frac{1}{  \frac{ - ( q_u^B - q_d^B ) }{M_1} v^{}_d
   \sqrt{ \frac{M_1^{\,2} }{  ( q_u^B - q_d^B )^2 } \frac{ v^2 }{ v_u^2 v_d^2 }+1 } }    \nonumber\\
 &=&     \frac{1}{ v_d \, \frac{ v }{v_u v_d}   } N            =  N\sin{\b}    \\
\left( O^{\chi}  \right)_{31}  &=&  \frac{1}{ \sqrt{ \frac{ M_1^{\,2} }{  ( q_u^B - q_d^B)^2} 
\frac{ v^2 }{ v_u^2 v_d^2 } + 1 } }          \nonumber\\  
  &=&     \frac{1}{  \frac{M_1}{-  (q_u^B - q_d^B) \, v^{}_u} \,\, v^{}_u  
\sqrt{ \frac{  (q_u^B - q_d^B)^2}{M_1^{\,2} } + 
          \frac{ v^2 }{v_u^2 v_d^2} }   }     = NQ_1 \cos{\b}  \\
\nonumber\\
\left( O^{\chi} \right)_{12}&=&  \frac{v^{}_u}{\sqrt {v^{\,2}_u + v^{\,2}_d} }   = \sin{\b}     \\
\left( O^{\chi} \right)_{22}&=&  \frac{v^{}_d}{\sqrt {v^{\,2}_u + v^{\,2}_d} }    =   \cos{\b}                                                     \\
\left( O^{\chi} \right)_{32}&=&    0         \label{coeffic1}    \\
\nonumber\\
\left( O^{\chi} \right)_{13}  &=&   \frac{1}{ \sqrt{ 1 + \frac{  ( q_u^B - q_d^B )^2  }{ M_1^{\,2} } 
    \frac{  v_u^{\,2} v_d^{\,2} }{  v^2  } } }   \left( 
 - \frac{  ( q_u^B - q_d^B ) }{ M_1 } \right)  \frac{v_u v_d^2}{  v^2 }  \nonumber\\
& =&   N   \left[  - \frac{ ( q_u^B - q_d^B ) }{ M_1 } v_u \cos\beta \right] \cos\beta  
 =  N {\overline Q}_1 \cos{\b}  \label{coeff_higgs_up}  \\
\left( O^{\chi}  \right)_{23} &=&  -   \frac{1}{ \sqrt{ 1 + \frac{  ( q_u^B - q_d^B )^2  }{ M_1^{\,2} } 
    \frac{  v_u^{\,2} v_d^{\,2} }{ v^2  } } }   \left( -
  \frac{  ( q_u^B - q_d^B ) }{ M_1 } \right) \frac{  v_u^2 v_d }{ v^2}       \nonumber\\
 &=&   -   N   \left[ 
  \frac{ -  ( q_u^B - q_d^B ) }{ M_1 } v_u \cos\beta  \right] \sin\beta 
 = -  N {\overline Q}_1  \sin\beta    \label{coeff_higgs_down}     \\
\left( O^{\chi}  \right)_{33}  &=&   \frac{1}{ \sqrt{ 1 + \frac{  ( q_u^B - q_d^B )^2  }{ M_1^{\,2} } 
    \frac{  v_u^{\,2} v_d^{\,2} }{  v^2 } }}  =   N.        \label{coeffic2}         
\eea
It can be easily checked that this is an orthogonal matrix
\bea
\left( O^{\chi}  \right)^{T}  O^{\chi} = {1}_{3\times 3}.
\eea

\subsection{ Appendix: Vanishing of the amplitude $\Delta^{\la \mu \nu}$ for  on-shell external physical states} 
An important property of the triangle amplitude is its vanishing for on-shell external physical states.  

The vanishing of the amplitude $\Delta$ for on-shell physical states can be verified once 
we have assumed conservation of the vector currents. This is a simple example 
of a result that, in general, goes under 
the name of the Landau-Yang theorem. In our case we use only the expression of the triangle in Rosenberg parametrization 
\cite{Rosenberg} and its gauge invariance to 
obtain this result. 
We stress this point here since if we modify the Ward identity on the correlator, as we are going to discuss next, 
additional interactions are needed in the analysis of processes mediated by this diagram in order to obtain consistency 
with the theorem. 

We introduce the 3 polarization four-vectors for the $\lambda$, $\mu$, and $\nu$ lines, denoted by ${\bf e}$,  
${\bf \epsilon_1}$ 
and ${\bf \epsilon_2}$ respectively, and we use the Sudakov parameterization of each of them, using the massless vectors 
$k_1$ and $k_2$ as a longitudinal basis on the light-cone, plus transversal $(\perp)$ components which are orthogonal to the 
longitudinal ones. We have 
\beqa
{\bf e} &=& \alpha (k_1 - k_2) + {\bf e_\perp} \qquad {\bf \varepsilon_1}= a k_1 + {\bf \varepsilon_{1\perp}}
 \qquad {\bf \varepsilon_2}= b k_2 + {\bf \varepsilon_{2\perp}}, \nonumber \\
\eeqa
where we have used the condition of transversality ${\bf e}\cdot k=0, {\bf \varepsilon_1}\cdot k_1=0, 
{\bf \varepsilon_2}\cdot k_2=0 $, the external lines being now physical.  Clearly ${\bf e_\perp}\cdot k_1={\bf e_\perp}\cdot k_2=0$, and similar relations hold also for ${\bf \varepsilon_{1 \perp}}$ and ${\bf \varepsilon_{2 \perp}}$, all 
the transverse polarization vectors being orthogonal to the light-cone spanned by $k_1$ and $k_2$. From gauge invariance 
on the ${\mu \nu}$ lines in the invariant amplitude, we are allowed to drop the light-cone components of the polarizators for 
these two lines 
\beq
\Delta^{\lambda \mu\nu}{\bf e}_\lambda {\bf \varepsilon_{1\mu}}{\bf \varepsilon_{2\nu}}= 
\Delta^{\lambda \mu\nu}{\bf e}_\lambda {\bf \varepsilon_{1\mu\perp}}{\bf \varepsilon_{2\nu\perp}}, 
\eeq
and a simple computation then gives (introducing ${\bf e}_\perp\equiv (0, \vec{{\bf e}})$ and similar)
\beqa
\Delta^{\lambda \mu\nu}{\bf e}_\lambda {\bf \varepsilon_{1\mu\perp}}{\bf \varepsilon_{2\nu\perp}} &=&
\underline{a}_1 \epsilon[k_1 - k_2,{\bf \varepsilon_{1\perp}},{\bf \varepsilon_{2\perp}},{\bf e}]
=\underline{a}_1 \epsilon[k_1 - k_2,{\bf \varepsilon_{1\perp}},{\bf \varepsilon_{2\perp}},
\alpha (k_1 - k_2) + {\bf e_\perp} ] \nonumber \\
&\propto & \left(\vec{\bf \varepsilon}_{1\perp}\times  \vec{\bf \varepsilon}_{2\perp}\right)\cdot \vec{\bf e}_\perp =0,   
\eeqa
since the three transverse polarizations are linearly dependent. Notice that 
this proof shows that $Z \to \gamma \gamma$ with all three particles on-shell 
does not occur. As usual one needs extreme care when 
massless 
fermions are running in the loop. The situation is analogous to that encountered in 
spin physics in the analysis of the EMC result, 
where the puzzle was resolved \cite{CollinsMueller} by moving to the massless fermion case starting from 
off-mass shell external lines.

\section{Appendix. Massive versus massless contributions}
Here we briefly discuss the computation of the mass contributions to the 
amplitude. We start from the massless fermion limit. 
The anomaly coefficient in rel.~(\ref{anom_coeff}) can be obtained starting from the triangle diagram 
in momentum space. For instance we get
\beqn
&&\Delta^{\lambda \mu \nu, ij}_{BSU(2)SU(2)}   
=   g^{}_{B} g^{\,2}_{2} \,  Tr[\tau^{i} \tau^{j}]  \sum_{f} q^{fL}_{B} {\bf \Delta}^{L \lambda \mu \nu}  \nonumber\\
&=& g^{}_{B} g^{\,2}_{2} \,  Tr[\tau^{i} \tau^{j}]  \sum_{f} q^{fL}_{B} 
 (i)^{3} \int \frac{d^{4} q}{ (2 \pi)^{4} } 
\frac{Tr[ \gamma^{\lambda} P_{L} ( \ds{q} - \ds{k} ) \gamma^{\nu} P_{L} (\ds{q} - \ds{k}_{1} ) 
\gamma^{\mu} P_{L} \ds{q} ]}{ q^2 (q -k_1)^2 (q-k)^2 }     \nonumber\\
&&+ (k_1 \rightarrow k_2, \mu \rightarrow \nu)     \nonumber\\
&=&  g^{}_{B} g^{\,2}_{2} \,  Tr[\tau^{i} \tau^{j}] \frac{1}{8} \sum_{f} q^{fL}_{B} 
 (i)^{3} \int \frac{d^{4} q}{ (2 \pi)^{4} } 
\frac{Tr[ \gamma^{\lambda}  (1 - \gamma^{5}) ( \ds{q} - \ds{k} ) \gamma^{\nu}  (1-\gamma^{5}) 
 (\ds{q} - \ds{k}_{1} ) 
\gamma^{\mu} (1-\gamma^{5})  \ds{q} ]}{ q^2 (q -k_1)^2 (q-k)^2 }     \nonumber\\
&&+ (k_1 \rightarrow k_2, \mu \rightarrow \nu)     \nonumber\\
\eeqn  
and isolating the four anomalous contributions of the form $\bf AAA$, $\bf AVV$, $\bf VAV$ 
and $\bf VVA$ we obtain

\beq
D^{L}_{B} = \frac{1}{8} Tr[q^{fL}_{B}] \equiv - \frac{1}{8} \sum_{f} q^{fL}_{B}.
\eeq
Similarly we obtain
\beqn
&&\Delta^{\lambda \mu \nu}_{BBB}  
=   g^{\,3}_{B}  \, \sum_{f} (q^{fR}_{B})^{3} {\bf \Delta}^{R \lambda \mu \nu}  
+  g^{\,3}_{B}  \, \sum_{f} (q^{fL}_{B})^{3} {\bf \Delta}^{L \lambda \mu \nu} \nonumber\\
&=& g^{\,3}_{B}  \, \sum_{f} (q^{fR}_{B})^{3} 
 (i)^{3} \int \frac{d^{4} q}{ (2 \pi)^{4} } 
\frac{Tr[ \gamma^{\lambda} P_{R} ( \ds{q} - \ds{k} ) \gamma^{\nu} P_{R} (\ds{q} - \ds{k}_{1} ) 
\gamma^{\mu} P_{R} \ds{q} ]}{ q^2 (q -k_1)^2 (q-k)^2 }      \nonumber\\
&&+ \, g^{\,3}_{B}  \, \sum_{f} (q^{fL}_{B})^{3} 
 (i)^{3} \int \frac{d^{4} q}{ (2 \pi)^{4} } 
\frac{Tr[ \gamma^{\lambda} P_{L} ( \ds{q} - \ds{k} ) \gamma^{\nu} P_{L} (\ds{q} - \ds{k}_{1} ) 
\gamma^{\mu} P_{L} \ds{q} ]}{ q^2 (q -k_1)^2 (q-k)^2 }    \nonumber\\
&&\,+ \,\, (k_1 \rightarrow k_2, \mu \rightarrow \nu)     \nonumber\\
&=&  g^{\,3}_{B}  \,  \frac{1}{8} \sum_{f} (q^{fR}_{B})^{3} 
 (i)^{3} \int \frac{d^{4} q}{ (2 \pi)^{4} } 
\frac{Tr[ \gamma^{\lambda}  (1 + \gamma^{5}) ( \ds{q} - \ds{k} ) \gamma^{\nu}  (1+\gamma^{5}) 
 (\ds{q} - \ds{k}_{1} ) 
\gamma^{\mu} (1 + \gamma^{5})  \ds{q} ]}{ q^2 (q -k_1)^2 (q-k)^2 }     \nonumber\\
&&+ \,  g^{\,3}_{B}  \,  \frac{1}{8} \sum_{f} (q^{fL}_{B})^{3} 
 (i)^{3} \int \frac{d^{4} q}{ (2 \pi)^{4} } 
\frac{Tr[ \gamma^{\lambda}  (1 - \gamma^{5}) ( \ds{q} - \ds{k} ) \gamma^{\nu}  (1 - \gamma^{5}) 
 (\ds{q} - \ds{k}_{1} ) 
\gamma^{\mu} (1 - \gamma^{5})  \ds{q} ]}{ q^2 (q -k_1)^2 (q-k)^2 }     \nonumber\\
&&+ (k_1 \rightarrow k_2, \mu \rightarrow \nu)     \nonumber\\
\eeqn

\beqa
&& D^{L}_{B} = \frac{1}{8} Tr[q^{fL}_{B}] \equiv - \frac{1}{8} \sum_{f} q^{fL}_{B} \nonumber\\
D^{}_{BBB}&=&   \frac{1}{8} Tr[q^{3}_{B}] = \frac{1}{8} \sum_{f} \left[ (q^{fR}_{B})^{3}
 - (q^{fL}_{B})^{3}  \right]. \nonumber 
\eeqa

the other coefficients reported in eq. (\ref{DDD}) are obtained similarly.

\section{Appendix. CS and GS terms rotated}

The rotation of the CS and the GS terms into the physical fields and the goldstone gives
\ba
&&V^{BYY}_{CS}=d_1 \langle B Y\wedge F_Y\rangle=(-i)d_1\varepsilon^{\lambda\mu\nu\alpha}(k_{1\alpha}-k_{2\alpha})
\left[(O^{A\,T})_{21}^{2}(O^{A\,T})_{32}\right]Z^{\lambda} A_{\g}^{\mu}A_{\g}^{\nu}+\dots
\nonumber\\
&&V^{BWW}_{CS}=c_1 \langle \varepsilon^{\mu\nu\rho\sigma} B_{\mu}C_{\nu\rho\sigma}^{Abelian}\rangle=
(-i)c_1\varepsilon^{\lambda\mu\nu\alpha}(k_{1\alpha}-k_{2\alpha})
\left[(O^{A\,T})_{11}^{2}(O^{A\,T})_{32}\right]Z^{\lambda} A_{\g}^{\mu}A_{\g}^{\nu}+\dots
\nonumber\\
&&V^{bYY}_{GS}=\frac{C_{YY}}{M}b F_Y\wedge F_Y=4 \frac{C_{YY}}{M}b \varepsilon^{\mu\nu\rho\sigma}k_{\mu}k_{\nu}Y_{\rho}Y_{\sigma}=
4 \frac{C_{YY}}{M}\varepsilon^{\mu\nu\rho\sigma}k_{\mu}k_{\nu}\left[
O^{\chi}_{31} (O^{A\,T})_{21}^{2}\,\chi A_{\g}^{\mu}A_{\g}^{\nu}
\right.\nonumber\\
&&\hspace{5cm}\left.
+(O^{\chi}_{32}C_1+O^{\chi}_{33}C_1^{\prime}) (O^{A\,T})_{21}^{2}G_Z A_{\g}^{\mu}A_{\g}^{\nu}\right]
+\dots\nonumber\\
&&V^{bWW}_{GS}=\frac{F}{M}b Tr\left[F_W\wedge F_W\right]
=4 \frac{C_{YY}}{M}\frac{b}{2} \varepsilon^{\mu\nu\rho\sigma}k_{\mu}k_{\nu}W^{i}_{\rho}W^{i}_{\sigma}=
4 \frac{F}{M}\varepsilon^{\mu\nu\rho\sigma}k_{\mu}k_{\nu}\left[
O^{\chi}_{31} (O^{A\,T})_{11}^{2} \,\chi A_{\g}^{\mu}A_{\g}^{\nu}
\right.\nonumber\\
&&\hspace{5cm}\left.
+(O^{\chi}_{32}C_1+O^{\chi}_{33}C_1^{\prime}) (O^{A\,T})_{11}^{2}\right]G_Z A_{\g}^{\mu}A_{\g}^{\nu}
+\dots\nonumber\\
\ea
These vertices appear in the cancelation of the gauge dependence in s-channel exchanges of Z gauge bosons in the $R_\xi$ gauge. The dots refer to the additional contributions, proportional to interactions of $\chi$, the axi-Higgs, with
the neutral gauge bosons of the model.

\end{document}